À Vinko Globokar, le maestro

# "Spectral Harmony"

# Towards the Unification of Timbre and Speech: Resolution of the Young Mahler-Schönberg Dream

Metric Space Models for Music on Informational Geometry 2

Yusei TAMURA (1, Shigekazu ISHIHARA (2 and Ken ITO (1

(1 The University of Tokyo

(2 Hiroshima International University

In this article, we present a method of spectral serialization—envisioned by G. Mahler and A. Schönberg [1] between the late 19th and early 20th centuries—that enables the "selection of timbres rooted in musical structure." Based on information geometry, we evaluate the Wasserstein distance between polyphonic timbre spectra, as demonstrated in our previous paper [2], in terms of the second moment of spectral chords comprising three or more notes, expressed in squared hertz ($Hz^2$). By embedding these quantities in a Banach space, we analyze specific passages from works by R. Strauss, Mahler, A. Berg, and A. Webern from the perspective of timbre, showing how the same chord exhibits different affinities and divergences depending on orchestration. Taking the wah-wah mute as an example, we propose a model that treats instrumental timbre and speech within the same Banach space, and we attempt to quantify the "vowel triangle" in Hz. These efforts constitute a partial but integrated solution to the historical "three problems of Schönberg"—serialism, *Klangfarbenmelodie*, and *Sprechgesang*—which neither Mahler nor Schönberg in their lifetimes, nor the postwar generation of composers including Nono, Boulez, and Stockhausen, was able to resolve fully. We aim to reconsider musical thought as a whole from the perspective of timbre and to open new possibilities for interpretation, performance, and composition. All tools are publicly available on GitHub, so that anyone can reproduce and extend these analyses.

============================================================

1) 7-3-1 Hongo, Bunkyo Ward, Tokyo, JAPAN

2) 555-36 Kurose Gakuen-dai, Higashi-Hiroshima, Hiroshima JAPAN

Contact address:

À Vinko Globokar, le maestro

## 0:Introduction: Contrasting Perspectives on 'Timbre' — Ray Kurzweil and Luigi Nono

Ray Kurzweil is widely known today for his concept of the 'Singularity' [3], which he introduced in 2005. Yet his earlier work, dating from the 1970s, was concerned with a very different problem: shortly after graduating from the Massachusetts Institute of Technology, Kurzweil began developing a device that could read printed text aloud for people with visual impairments. In collaboration with AT&T Bell Laboratories, he subsequently developed and commercialized the world's first practical print-to-speech system, combining optical character recognition (OCR) with speech synthesis. The visually impaired musician Stevie Wonder was among its first users.

In 1982, while visiting Wonder's studio, Kurzweil was asked whether a computer could reproduce the sound of a real acoustic instrument.

Kurzweil subsequently appointed Wonder as an adviser and began developing a sampling-based synthesizer. In 1984, he introduced the Kurzweil K250, an instrument that enabled users to digitally record the sounds of acoustic instruments and reproduce and perform them through a keyboard or similar interface. The K250 thus represented an important step toward the digital reproduction of instrumental timbres that had previously been associated almost exclusively with acoustic sound production.

What Stevie Wonder sought was the ability to assign the sounds of real instruments—or digitally recorded samples sufficiently close to those sounds—to an electronic keyboard, which was considerably more convenient for him to play given his visual impairment [4].

Until the early 1980s, electronic and electroacoustic sounds were generally readily distinguishable from those produced by acoustic instruments. Electric pianos, electronic organs, and analog synthesizers all possessed characteristic sonic qualities of their own. The music of the West German group Kraftwerk, formed in 1970, provides a particularly clear example of a musical aesthetic in which the artificiality of electronic sound was not concealed but deliberately foregrounded. Electronic instruments were therefore often employed precisely in contexts that emphasized their difference from acoustic instruments.

The work of Stevie Wonder and Ray Kurzweil can be understood in this context as part of a broader technological and musical shift toward the reproduction of familiar acoustic sounds by electronic means. Rather than celebrating the sonic distinctiveness of electronic instruments, their work pursued the opposite goal: making electronically generated or reproduced sounds increasingly difficult to distinguish from their acoustic counterparts. In this sense, the development of sampling technology marked a transition from the production of explicitly 'synthetic' timbres toward the digital simulation of natural instrumental sounds.

This issue was particularly significant in the development of text-to-speech technologies for people with visual impairments. Here, the problem was not merely to reproduce speech intelligibly, but to reproduce the expressive and affective qualities of human speech. Although text-to-speech systems capable of conveying relatively sophisticated emotional expression did not become commercially widespread until the 2020s, the fundamental problem of how electronically generated sound can reproduce the expressive characteristics of natural human sound remains far from fully resolved.

Now, whilst 'instrumental sound' has been actively re-engaged with by computer-driven electronic media since the 1980s, composers such as Luigi Nono criticized this trend in the late 1980s, stating that 'the exploration of essential timbre had ceased, and efforts were confined to the superficial imitation of existing instruments [5].

There is no doubt that the innovation commissioned from Kurzweil by Stevie Wonder was, in this sense, a typically 'music-structure-independent' application. What, then, might be the nature of the 'conception of timbre directly linked to musical structure' that Schönberg proposed and which Nono and others sought to carry forward?

In our previous article, we outlined the key points of an approach from 'information geometry' that addresses this question. In this article, through a 'timbral-harmonic analysis' of the works of Gustav Mahler—a composer and conductor who, alongside Schönberg, envisioned the future of music through 'timbre' but passed away before realizing his full potential—we wish to demonstrate, using Wasserstein geometry, the realization of the 'serialization unifying timbre and speech=singing'—a concept imagined in the 1950s but never realized—as an example of the 'treatment of timbre directly linked to musical structure'.

## 1. Gustav Mahler's Prophetic Vision and his Untimely Death

According to the violist Natalie Bauer-Lechner (1858–1921) [6], a close friend of Gustav Mahler, Mahler articulated clear and distinctive views on instrumental timbre during his lifetime.

For Mahler, timbre was not merely a means of providing 'coloring of sound for effect'; rather, it served to differentiate melodic lines and clarify the structure of polyphonic textures. He also observed that when instruments are played at the extremes of their registers, or when unconventional playing techniques are employed, distinctive and often highly charged timbral qualities emerge. In this respect, his thinking anticipated several concerns that would become prominent in twentieth- and twenty-first-century music, including register-based listening, extended playing techniques, and the incorporation of noise as a musical element.

Mahler's observations also point toward a conception of timbre that encompasses spatial perception. He noted that the timbre of a *banda*—a group of brass instruments positioned behind or off the main stage, for example—could introduce a new dimension into musical experience: one that was not simply a matter of increased volume, but also of a ‘sense of distance’ and ‘spatial expansiveness’ [4]. These observations suggest that, for Mahler, timbre could function simultaneously as a means of articulating musical structure and as a means of shaping the listener’s perception of musical space.

Let us examine these ideas through several examples. In the present study, we focus on the brass instruments, and particularly on the trombone, in order to investigate the effects of timbral transformation from the perspective of mute usage.

Figure 1 shows the passage in the first movement of Mahler’s Symphony No. 3 (1895–96) in which muted trombones make their first appearance.

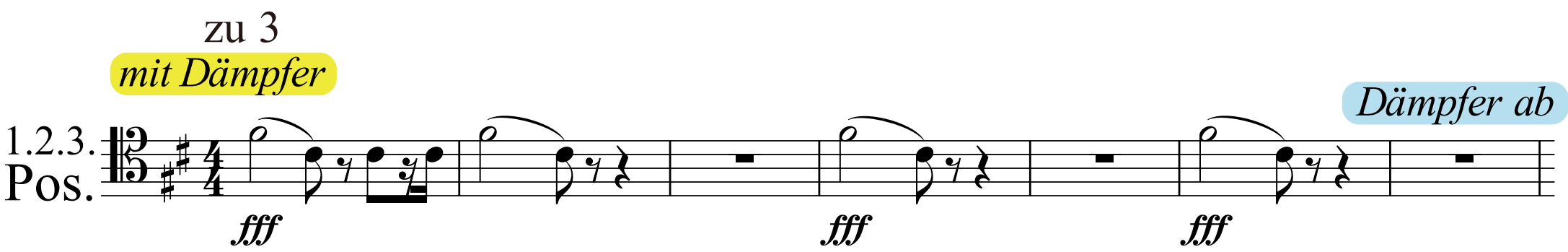

Fig. 1: G. Mahler Symphony 3, 1 mvt. Bar. 341～

The full score contains detailed performance instructions in Mahler’s own hand, including such directions as ‘attach the mute’ and ‘remove the mute’ (these instructions were, of course, subsequently copied into the individual orchestral parts by the copyists). As a conductor with extensive practical experience, Mahler was meticulous about such performance details. At this point in the movement, the three trombones play in unison at *ff* with mutes, causing their combined sound to stand out distinctly in terms of timbre.

Immediately before the B♭-major *ff* chord that follows, Mahler specifically instructs the players to remove their mutes [7]. The abrupt change in instrumentation thus produces a corresponding change in timbre at a structurally significant moment.

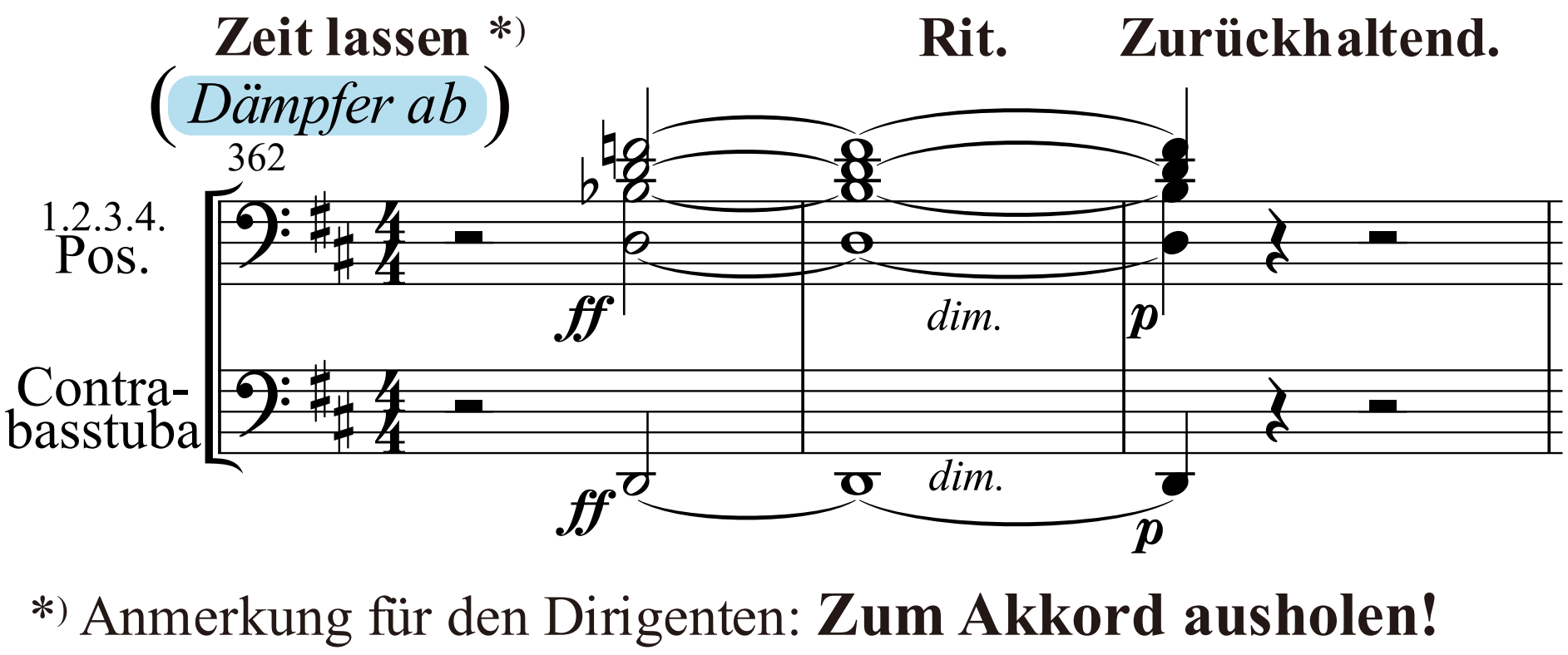

Fig. 2: G. Mahler Symphony 3, 1 mvt. Bar.362

The use of brass mutes took a significant new direction the following year, in 1897. Richard Strauss, who was also active as a conductor, commissioned instrument makers in Germany to produce high-quality mutes designed specifically to minimize, or even eliminate, intonation problems. He subsequently employed these mutes in *Don Quixote* (1897), using them in the brass section, including the tubas, in ways that altered the pitch relationships among the instruments.

The passage shown below is believed to be one of the earliest known examples in which all of the trombone parts are played with mutes while forming harmonies that are neither in unison nor separated by an octave [8].

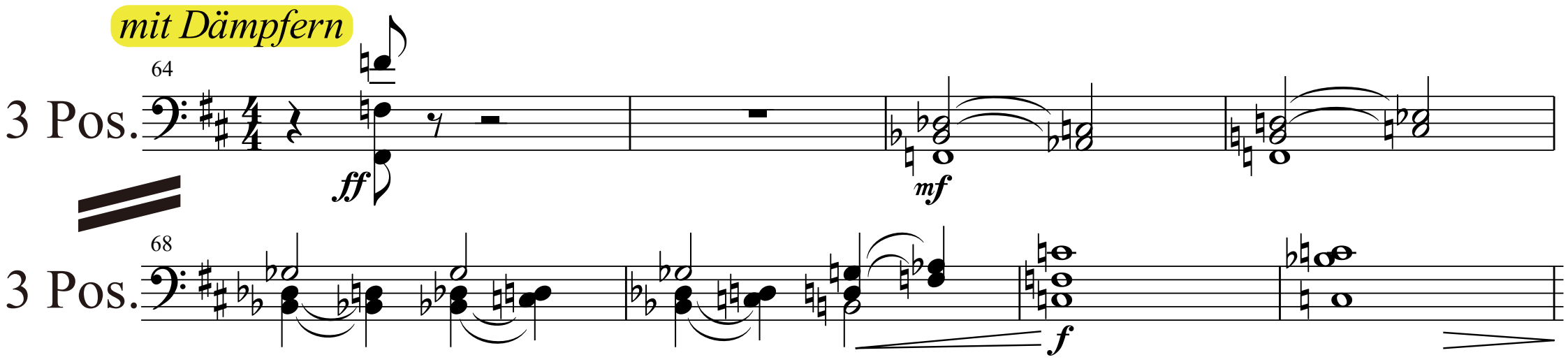

Fig. 3: R. Strauss “Don Quixote” Bar. 66～

It is a familiar experience among musicians that, in concert halls and rehearsal rooms alike, chords played by muted brass instruments—particularly their inner voices—can be more difficult to distinguish than the corresponding chords played without mutes. Following the pioneering example of R. Strauss described above, the use of mutes in brass sections gradually increased from the beginning of the twentieth century onward. In the early years, however, mutes were not widely available—as indicated by the fact that they were often custom-made—and their use consequently remained far from widespread [9].

It was in this historical context that Anton Webern composed his *Passacaglia*, Op. 1 (1908), which may be regarded as one of the earliest and most comprehensive examples of a work in which mutes are specified for all brass instruments throughout the entire composition. In what follows, we examine Webern’s use of mutes, focusing in particular on the trombone.

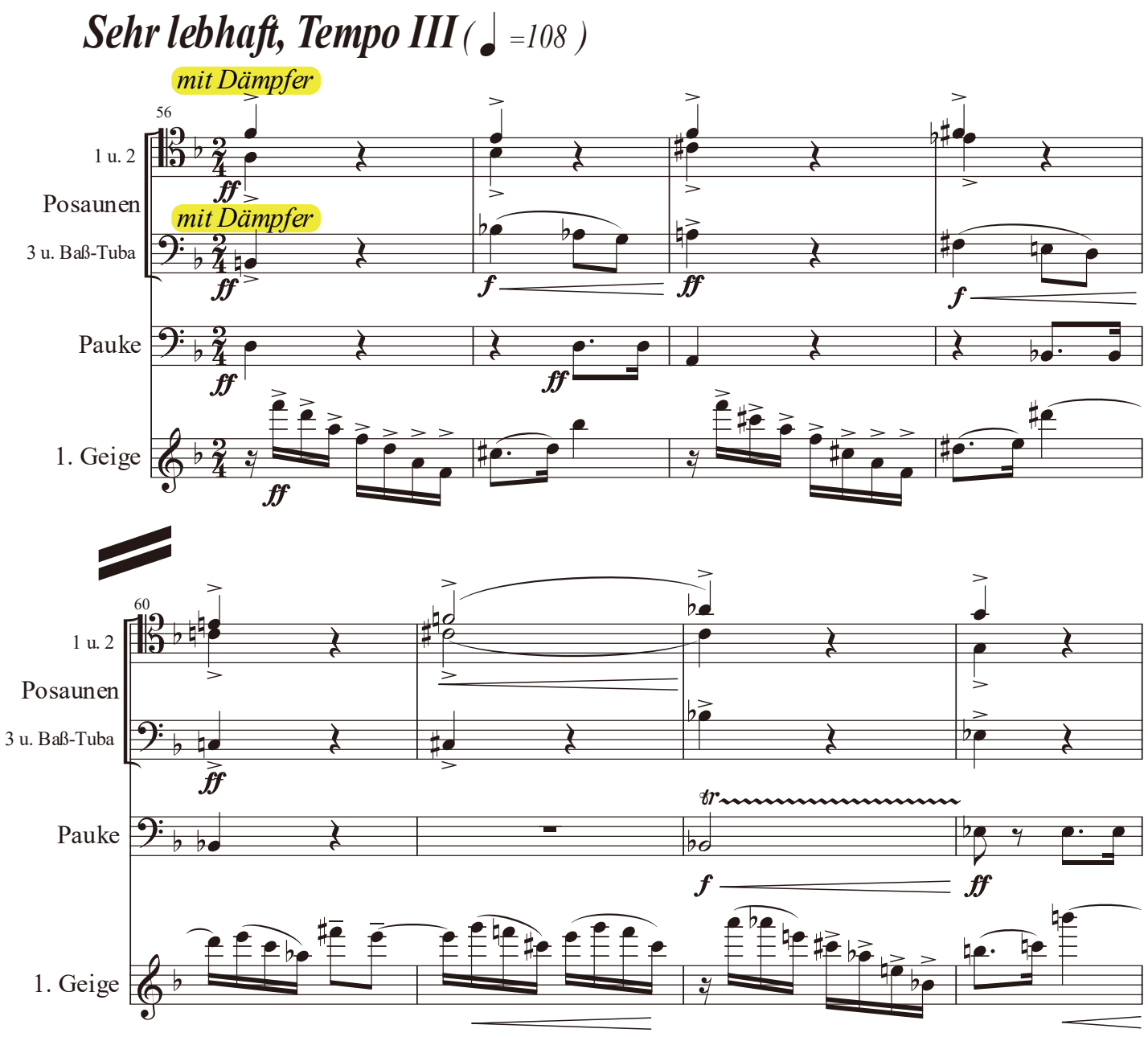

Fig. 4: Webern “Passacaglia” op.1 Bar.56～

Fig. 4, In bar 56, the trombone part forms part of a D-minor sonority articulated by the timpani on D, while the bass line of the section is on B♮. The resulting sonority may be understood as a kind of ‘third inversion’ of an added-sixth chord. Consequently, the harmony does not readily register as D minor; rather, it acquires an obscure and ambiguous quality, resembling a first-inversion diminished triad with an added seventh above B♮.

The use of mutes further contributes to the ambiguity of the overall trombone sonority, making its harmonic identity less clearly perceptible. The extent to which the muted sound obscures the harmonic structure will be examined in detail in the following section.

Arnold Schönberg, who conducted the premiere of this work with the Vienna Philharmonic Orchestra in November 1908, subsequently made extensive use of mutes in the brass instruments, particularly from 1909 onward. This practice is especially evident in his *Five Orchestral Pieces*, Op. 16, and most notably in the third movement, ‘Farben’ (‘Colours’).

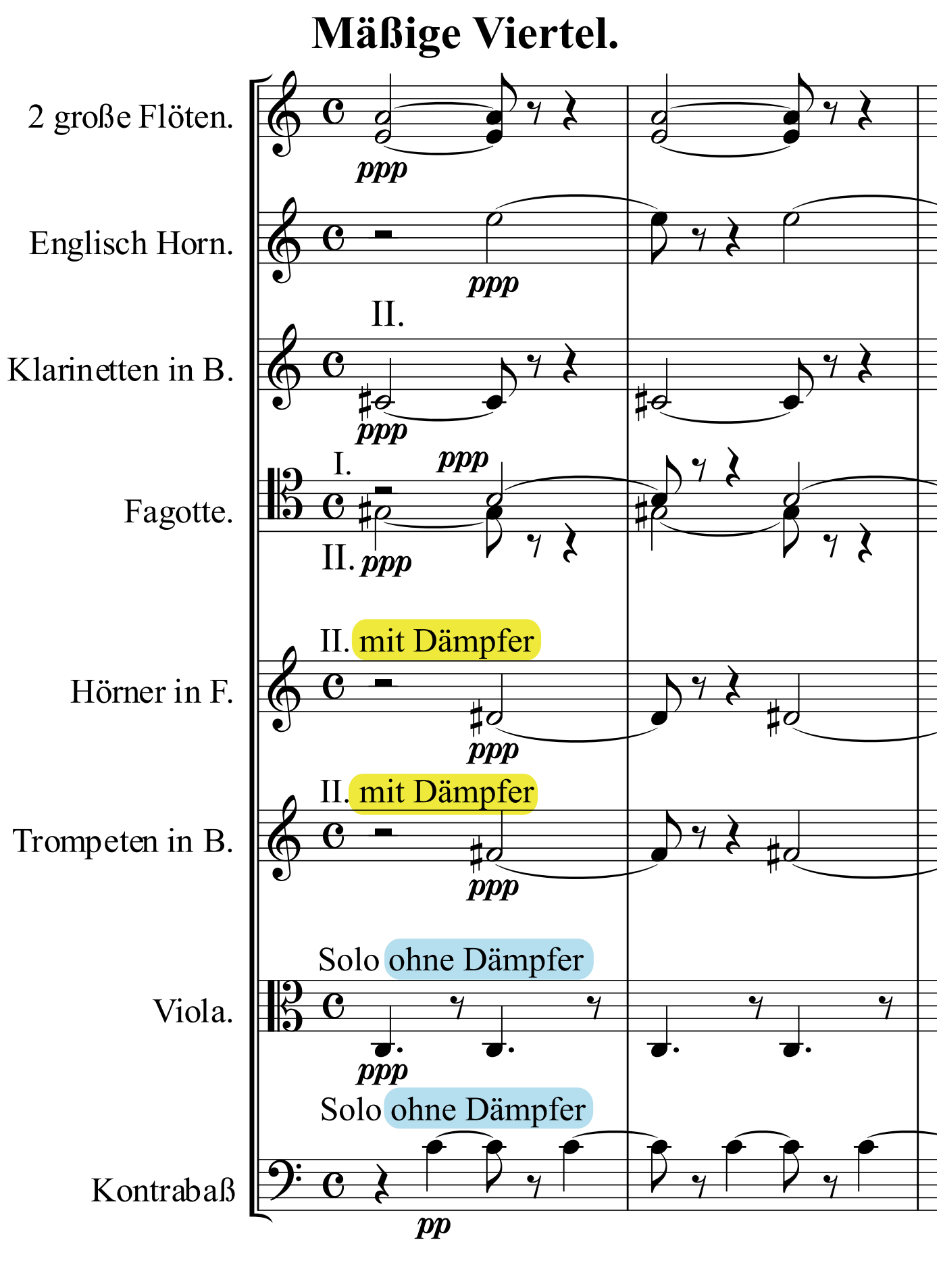

Fig. 5: Schönberg “Fünf Orchesterstücke” op.16, 3 “Farben”

However, more advanced examples can certainly be found in Webern's *Six Pieces for Orchestra*, Op. 6, composed in 1909. Not only are all the brass instruments played with mutes, but the work also contains examples of a 'set of mutes' combined with a wider variety of playing techniques, including glissandi on the trombone.

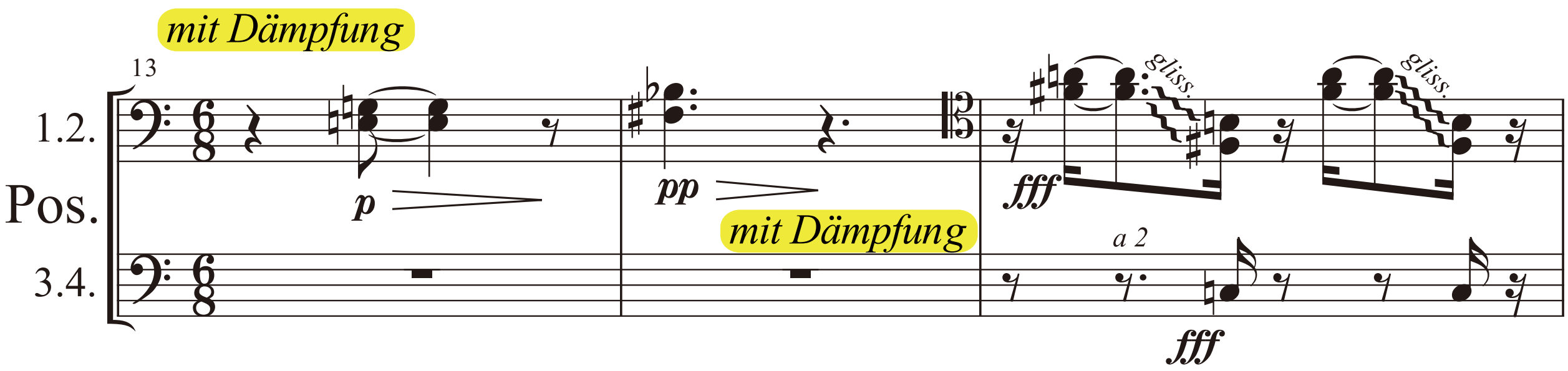

Fig. 6: Webern "Sechs Stücke für großes Orchester" op.6, 2 Bar. 13～

What, then, are the examples in Mahler's own late works that may have inspired this line of thought? To explore this question, let us turn to the first movement of his Ninth Symphony.

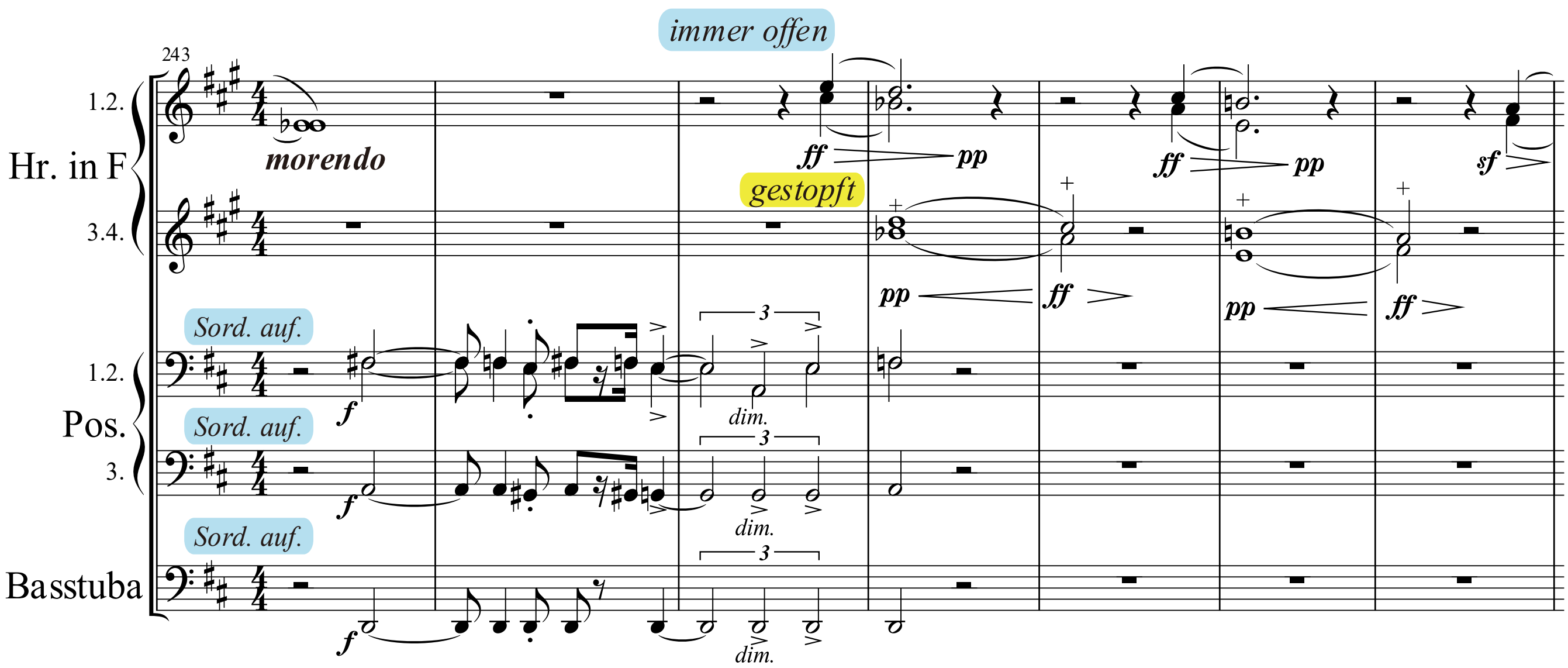

Fig. 7: Mahler Symphony 9 1. Mvt. Bar.243～

The chords played by the muted trombone section form both major and minor triads on D; however, the intervallic relationships are not always clearly perceptible. Rather, the major triad appears to transform into the minor triad through an obscure and gradual transition.

The open G and E♭ played by the horns, flanking F—the minor third of D minor—undergo a timbral transformation from open to stopped during the sustain, thereby 'resolving' onto the minor third D–F. A phenomenon resembling a partially 'multi-exposed' photograph thus emerges in the harmonic progression. At the same time, because the trombones are muted, the contrast between their sound and the open-pipe sound of the horns is softened. Moreover, as the timbre of the horns changes to *gestopft* ('stopped') midway through the sustained notes, the D-minor sonority produced by the trombones gradually disappears. Ultimately, after the interval E♭–G is transformed into D–F through the *gestopft* third, the horn timbre returns to its open-pipe sound.

Schönberg referred to this approach in Mahler's final symphony as a form of '**Klangfarbenmelodie**' [10] in his ***Harmonielehre*** (*Theory of Harmony*, 1911). Written shortly after Mahler's death, the treatise reflects Schönberg's recognition of the possibilities inherent in Mahler's treatment of timbre, as well as his sense of loss at Mahler's premature death. Mahler himself, however, was unable to pursue these possibilities any further.

## 2. Klangfarben-Harmonic Analysis of Spectral Geometry in Mahler and Webern

In the following, we attempt to reinterpret the musical context examined in the previous section quantitatively by means of spectral information geometry.

First, taking Mahler's *Symphony No. 3*, as shown in Fig. 1, we use the Wasserstein distance [11] to quantify the extent to which the F♯ played in unison by the three muted trombones differs spectrally from the same note played without mutes.

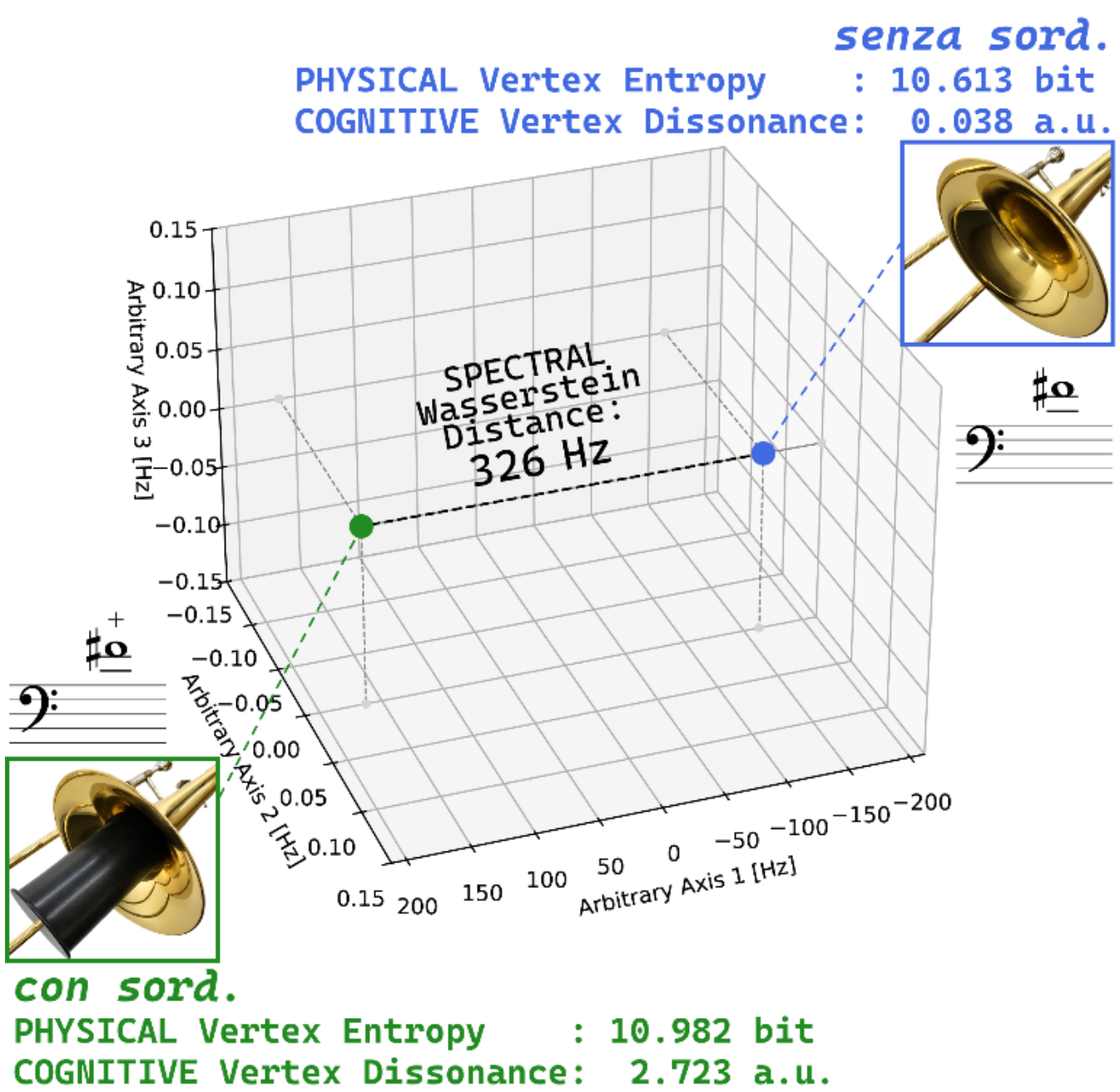

Fig. 1-b: Topological mapping of spectra: G. Mahler Symphony 3, 1 mvt. Bar. 341～

Tone to tone " F# " (●···vertex) , ------ : Wasserstein path

Assuming that the same pitch is produced by a sine wave, its frequency would be approximately 370 Hz. However, a pure sine wave cannot exist as a perfectly sustained physical sound in the external environment; any audible sound inevitably exhibits fluctuations in its temporal, spectral, and phase characteristics. The total spectral displacement produced by fitting a mute to the trombone is 326 Hz, which is of the same order of magnitude as the centre frequency itself. The resulting spectral change is therefore by no means negligible.

Furthermore, a spectrum in which all power is concentrated at a single frequency corresponds to a Dirac delta function, for which the Shannon entropy [12] is zero. Since real instrumental sounds contain information distributed across multiple spectral components, their Shannon entropy, when represented as physical sound spectra, can be expressed in bits.

At the same time, timbre is not merely a physical property of a sound spectrum but a perceptual phenomenon arising through auditory and cognitive processes. For this reason, we also evaluate sensory dissonance [14] using Sethares' Sensory Dissonance model [13]. For the normalized spectrum of the trombone F♯, the change in physical information content, as measured by the change in Shannon entropy, is relatively small: both the muted and unmuted sounds have entropy values of approximately 11 bits. By contrast, the corresponding values of Sethares' Sensory Dissonance differ by approximately a factor of 72, indicating a substantially greater sensory dissonance for the muted sound.

In other words, the comparison suggests that subjective sensory dissonance and physical information content, as measured by Shannon entropy, are only weakly correlated in this case.

We next examine how the change in interval between the F♯ and the subsequent C♯, as shown in the musical example from Mahler's *Symphony No. 3* in Fig. 1, affects the resulting spectrum.

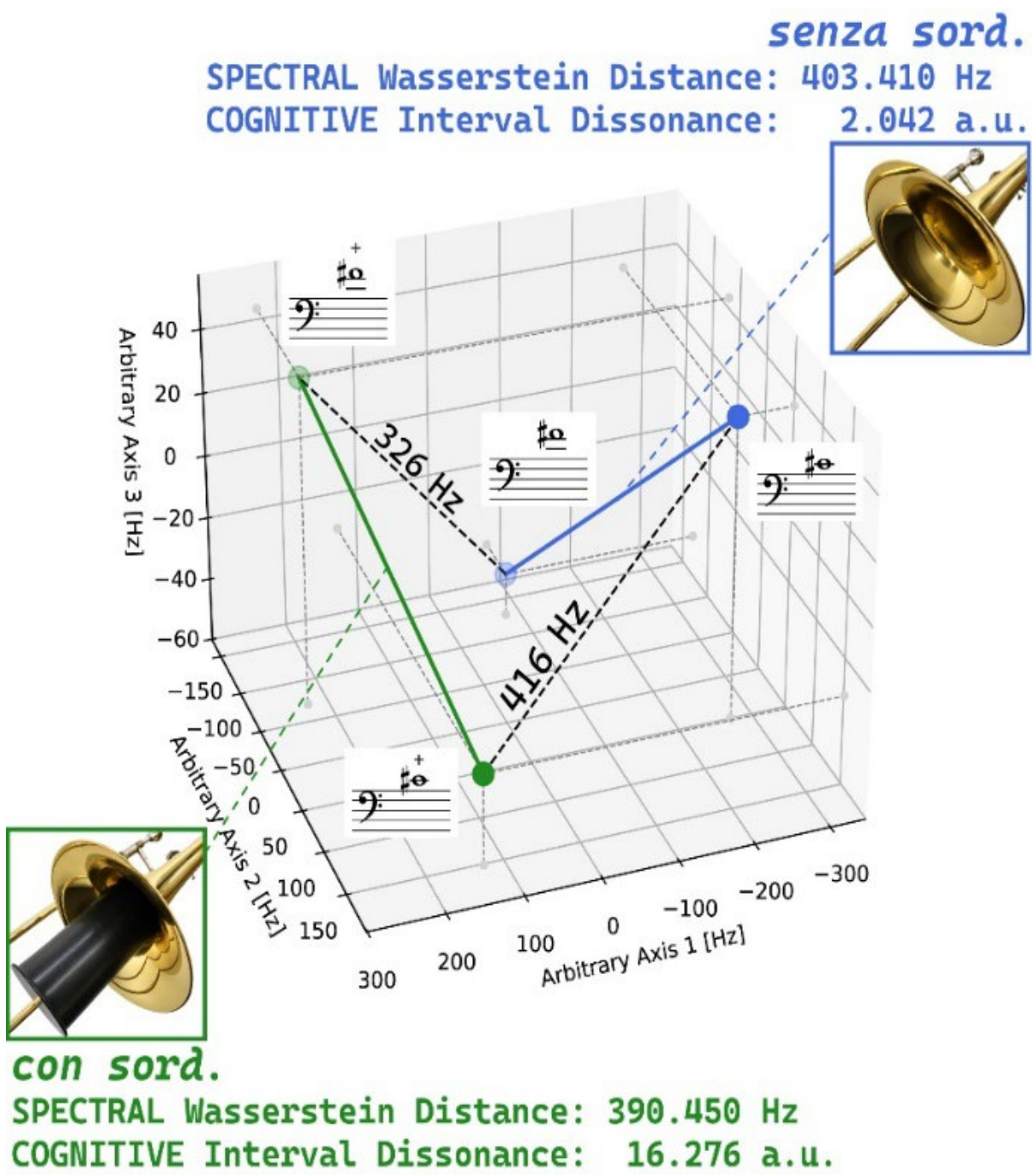

Fig. 1-b': Topological mapping of spectra: G. Mahler Symphony 3, 1 mvt. Bar. 341～:
Interval to interval "F# - C#" ——— : Simplex and ------- : Wasserstein path.

As mentioned earlier, the frequency of F♯ is approximately 370 Hz when represented as a sine wave, while that of C♯ is approximately 277 Hz. Therefore, if both tones were represented as sine waves, their Wasserstein distance would be approximately 93 Hz. In contrast, for the actual trombone sounds, the Wasserstein distance between the F♯ and C♯ is 403 Hz without a mute and 390 Hz with a mute—both substantially greater than the value obtained for the corresponding sine waves. This is to be expected, since actual instrumental timbres contain and therefore require the transport of many more spectral components than sine waves, which have no timbral complexity.

If we subtract the 93-Hz Wasserstein distance of the corresponding sine waves from each of these values, the resulting timbral gaps are 310 Hz and 297 Hz, respectively. Although the numerical values differ, they remain of the same order of magnitude.

A markedly different result emerges, however, when we consider the 'Sethares dissonance due to the difference in pitch' associated with the presence or absence of the trombone mute. According to Sethares' model, the interval between C♯ and F♯—a perfect fourth—produces a dissonance of approximately 2.0 arbitrary units when the trombone is played without a mute. By contrast, the same perfect fourth, when the trombone is played with a mute, produces a dissonance of approximately 16.3 arbitrary units. Thus, despite the interval being identical, the dissonance differs by a factor of approximately eight as a result of the change in timbre.

Returning to this perspective and reconsidering Fig. 1-b, we can see that the change in timbre produces a substantial change in the perceived dissonance of the interval, despite the pitch relationship remaining unchanged. This illustrates a fundamental point: 'even though the intervals are the same, they are not necessarily consonant to the same degree.' In other words, whereas conventional harmonic theory generally treats intervallic relationships independently of timbre, a timbre-sensitive approach to harmony reveals that the consonance or dissonance of an interval cannot always be determined from its pitch relationship alone. The present analysis therefore suggests that the consonance and dissonance of an interval should be evaluated in conjunction with the timbral characteristics of the sounds that constitute it.

Let us now apply this situation to polyphony involving three or more voices. Taking the section of Mahler's Symphony No. 3 shown in Fig. 2, where the three trombone parts play the chord B♭–D–F, we shall evaluate the distance between each pair of notes and the area of the triangle defined by them. As the Wasserstein distance satisfies the triangle inequality, it is always possible to define such a 'Wasserstein triangle' for three voices.

When four or more voices are involved, however, the corresponding set of pairwise Wasserstein distances cannot in general be represented by a single triangle [15].
Therefore, with a view to polyphonic generalization, we shall consider the Wasserstein distances $W_1(p_i, p_j)$, $W_1(p_j, p_k)$, $W_1(p_k, p_i)$ between the three timbre spectra $p_i, p_j, p_k$

$$I(\{p_i, p_j, p_k\}) = \frac{W_1(p_i, p_j)^2 + W_1(p_j, p_k)^2 + W_1(p_k, p_i)^2}{3} \cdots ①$$

This has been plotted on the graph as the Wasserstein second moment [16].

In what follows, we use Wasserstein geometry to evaluate the timbral affinity of chords [17].

However, when there are four or more voices, it is not always possible to define a triangle [15]. Therefore, with a view to polyphonic generalisation, we shall consider the Wasserstein distances $W_1(p_i, p_j)$, $W_1(p_j, p_k)$, $W_1(p_k, p_i)$ between the three timbre spectra $p_i, p_j, p_k$

$$I(\{p_i, p_j, p_k\}) = \frac{W_1(p_i, p_j)^2 + W_1(p_j, p_k)^2 + W_1(p_k, p_i)^2}{3} \cdots ①$$

This has been plotted on the graph as the Wasserstein second moment [16]. In what follows, we shall evaluate the timbral affinity of chords using Wasserstein geometry [17].

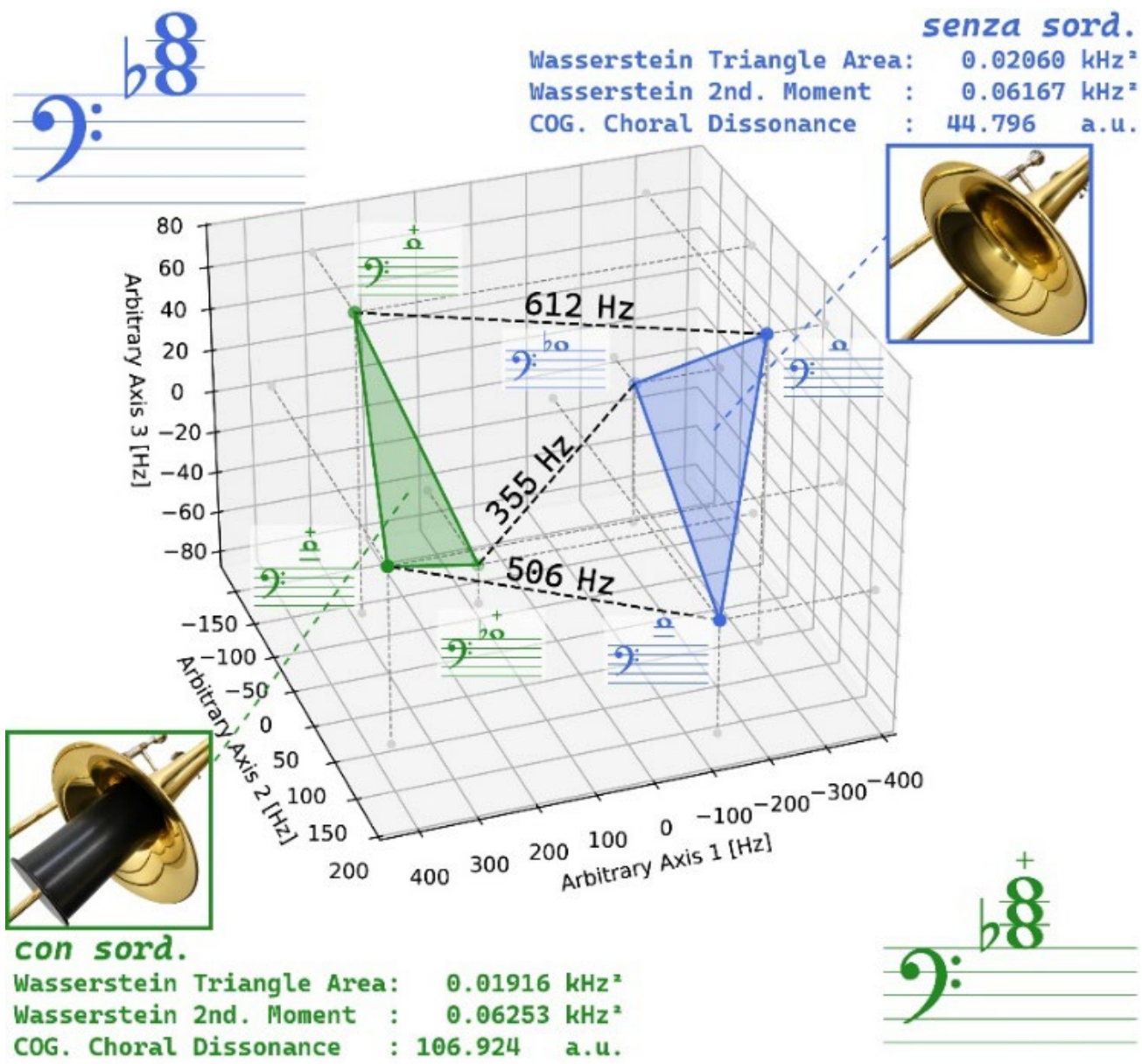

Fig. 2-b: Topological mapping of spectra : G. Mahler Symphony 3, 1 mvt. Bar. 362～:

Interval to interval " F# - C# " ——— : Simplex and ------- : Wasserstein path.

The frequencies of sine waves corresponding to the same pitches are approximately 233 Hz for B♭, 295 Hz for D, and 350 Hz for F. The Wasserstein distances between the corresponding vertex points for the muted and unmuted trombone sounds remain consistent with the previous example: despite the differences in pitch, their magnitudes are of the same order as the corresponding centre frequencies.

In contrast, a marked difference is observed in the area of the Wasserstein triangle. For the unmuted trombone, the area is 0.021 $kHz^2$, whereas with the mute fitted, it increases substantially to 0.030 $kHz^2$.

This difference becomes even more pronounced when the Wasserstein second moment is considered. The value increases from approximately 0.052 $kHz^2$ without a mute to 0.099 $kHz^2$ with a mute. The relationship between the area of the Wasserstein triangle and the Wasserstein second moment will be discussed in a later section.

Furthermore, the three-tone dissonance measured using Sethares' index increases by approximately a factor of three, from 12.5 to 34. These results demonstrate that, even when the harmonic structure represented on the grand staff is identical, both the physical affinity of timbral harmony—which takes the actual timbral characteristics of the sounds into account—and the corresponding perceptual dissonance can differ substantially.

Next, let us examine in a similar manner the 'change in timbre' of the horn in Mahler's *Symphony No. 9*, as shown in Fig. 7. The notes notated as B♭ and D for an F horn actually sound as E♭ and G, respectively. Expressed in terms of the frequencies of sine waves corresponding to these sounding pitches, they are approximately 311 Hz and 392 Hz, respectively. As in the case of the trombone, the Wasserstein distances between the corresponding vertices are of a similar order of magnitude.

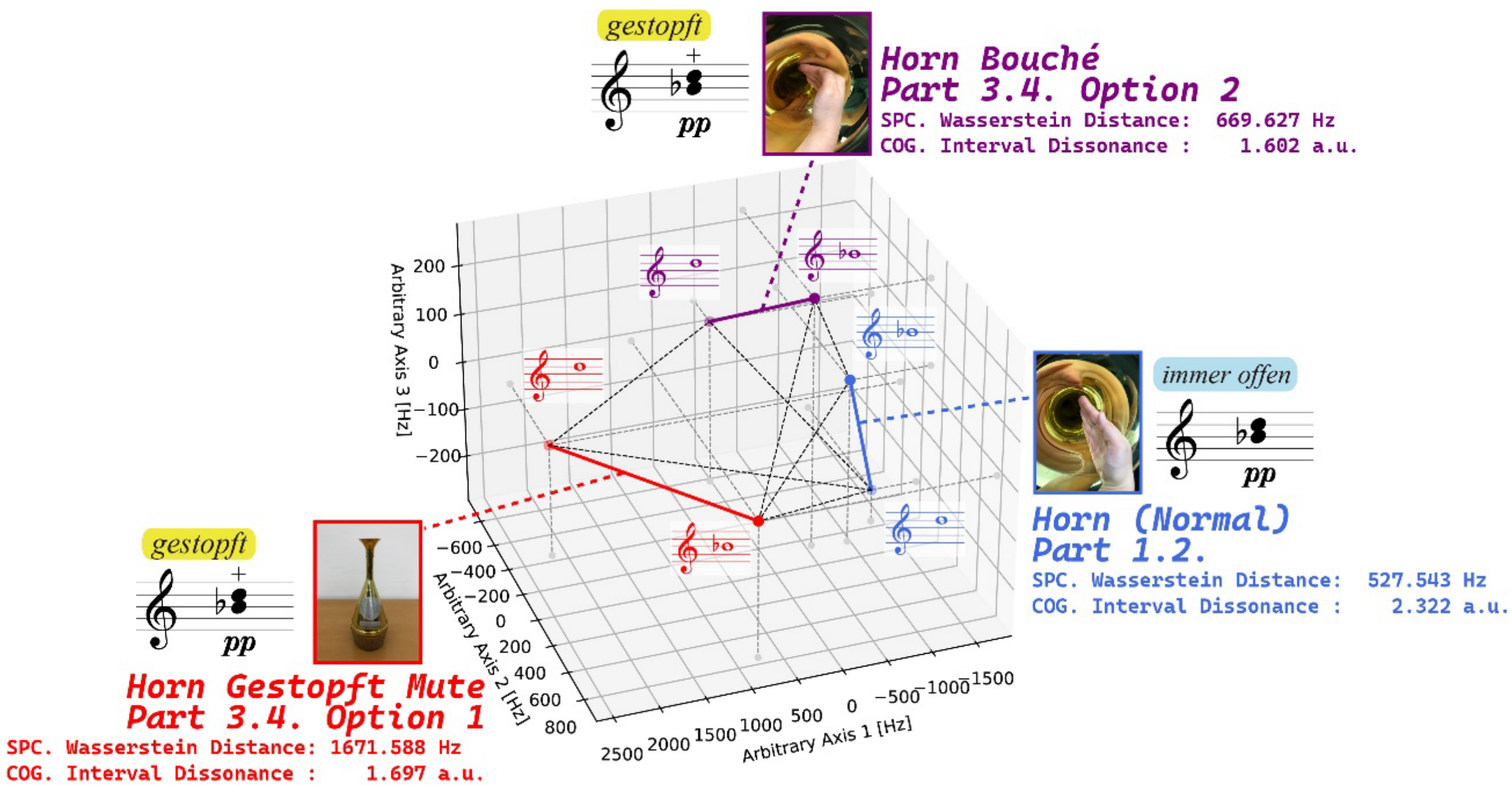

Fig. 7-b: Topological mapping of spectra: G. Mahler Symphony 9, 1 mvt. Bar. 246～ :
Horn " B♭ - D " Gestopft Mute Complex, Bouché-Complex & Ouvre-Complex

A comparison of standard horn-playing techniques with those used when a *bouchée* or trombone mute is employed, as indicated in Mahler's full score, is presented below.

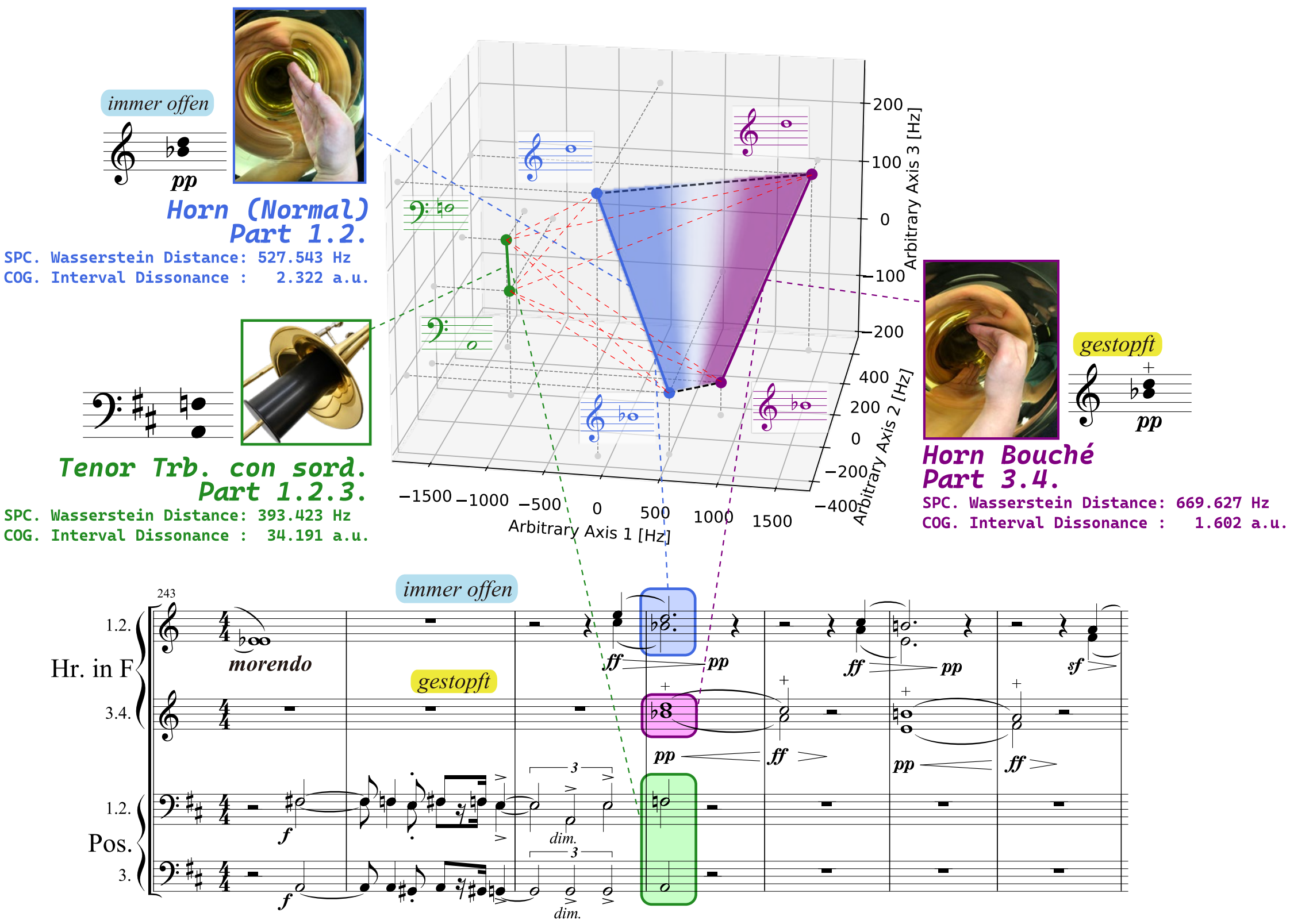

Fig. 7-c: Klangfarbenwechsel: G. Mahler Symphony 9, 1 mvt. Bar. 246～:

Horn “ B ♭ - D ” Ouvre-Complex → Bouché-Complex

With regard to the French horn, it is worth noting that, in contrast to the trombone, the degree of Sethares’ dissonance is significantly reduced when using articulation techniques. Compared to standard playing, this reduction is approximately 70 per cent with the ‘bouché’ technique and approximately 73 per cent when a ‘Gestopft’ mute is used. It is thought that this is related to the fact that the ‘bouché’ technique—which involves inserting the human hand, which is softer than a mute—acts as a low-pass filter, cutting out high-frequency components; we are continuing to analyse this phenomenon.

Consequently, as the sound of a trombone fitted with a mute differs significantly from both the timbres of a horn before and after the insertion of the hand, it has been quantitatively demonstrated that, even when multiple timbral groups resonate simultaneously, timbral masking is unlikely to occur and the polyphony is perceived as distinct. Whilst these were orchestrated based on what Gustav Mahler had empirically learnt and judged through his own ears, the method of Wasserstein geometry offers the significant advantage of being able to demonstrate such timbral alternations in a general and quantitative manner, including cases where such differences might be difficult to perceive intuitively.

Fig. 7-c illustrates the gradation of timbre in two dimensions to aid the reader’s understanding. However, in reality, the situation is as shown in Fig. 7-d below.

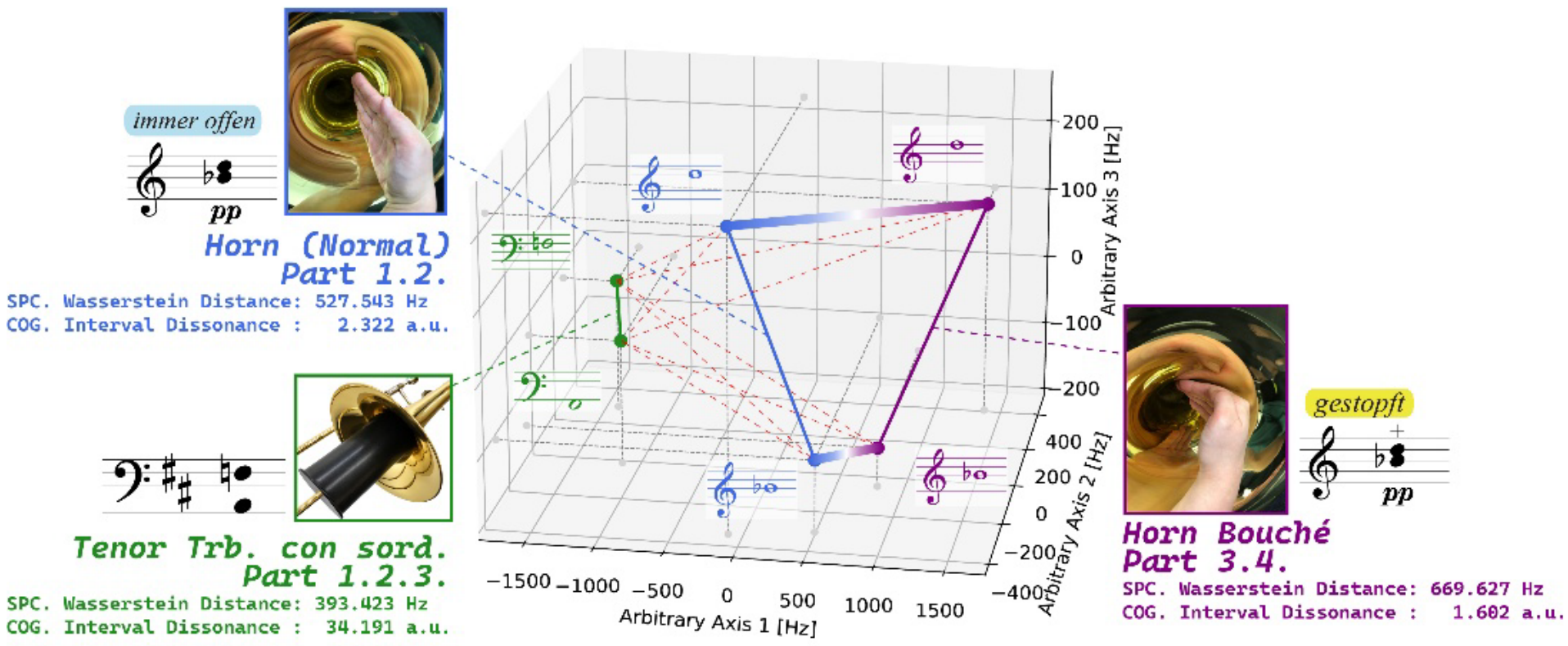

Fig. 7-d: Klangfarbenwechsel: Parallel

Although other forms of tonal progression, such as the example shown in Fig. 7-e, are possible, few further elaborations of this kind are found in the scores Mahler left during his lifetime.

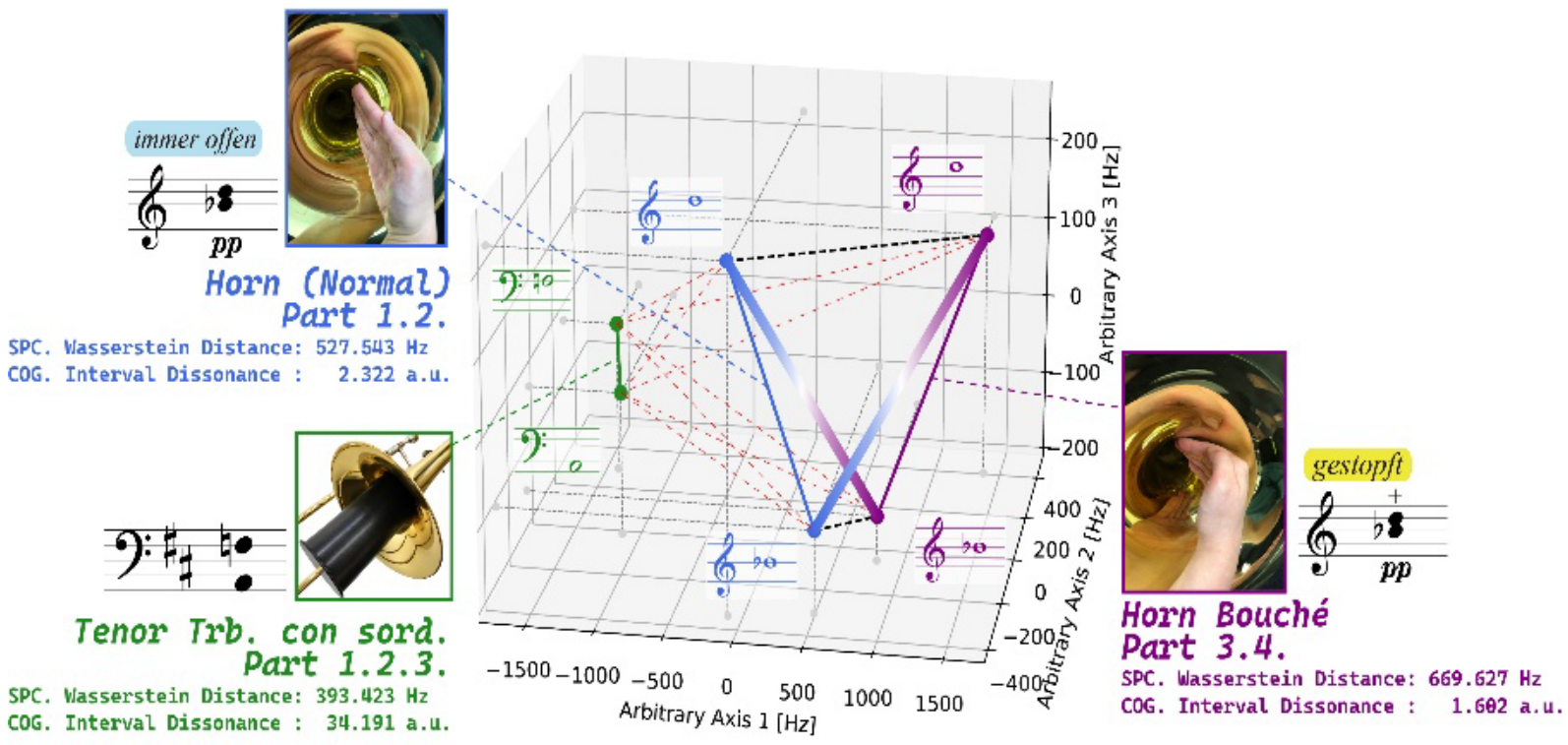

Fig. 7-e: Klangfarbenwechsel: Intersection

Fig. 8 shows the 'Invention on the Note H' from Act III, Scene 2 of Alban Berg's (1885–1935) *Wozzeck*. This provides a clear example of *Klangfarbenmelodie* in the strict sense of the term. By contrast, the pointillistic treatment of timbre developed by Webern from the 1920s onward often involved simultaneous or successive changes in pitch, and therefore cannot be regarded as an expression based exclusively on timbre.

Conceived in 1914, following Mahler's death, and provisionally completed in 1917 during the First World War, *Wozzeck* was not fully orchestrated until 1922. The first concert performance of excerpts from the opera took place in 1924, while the complete work was not premiered until 1925. Performances in Germany were subsequently prohibited following the Nazi seizure of power.

The musical form envisioned by the young Mahler and Schönberg—in which musical development proceeds essentially through timbre alone, without changes in pitch—had to await the period after the Second World War for its more systematic development in the works of composers such as Pierre Boulez and Karlheinz Stockhausen, who also incorporated electronic media into their compositional practice.

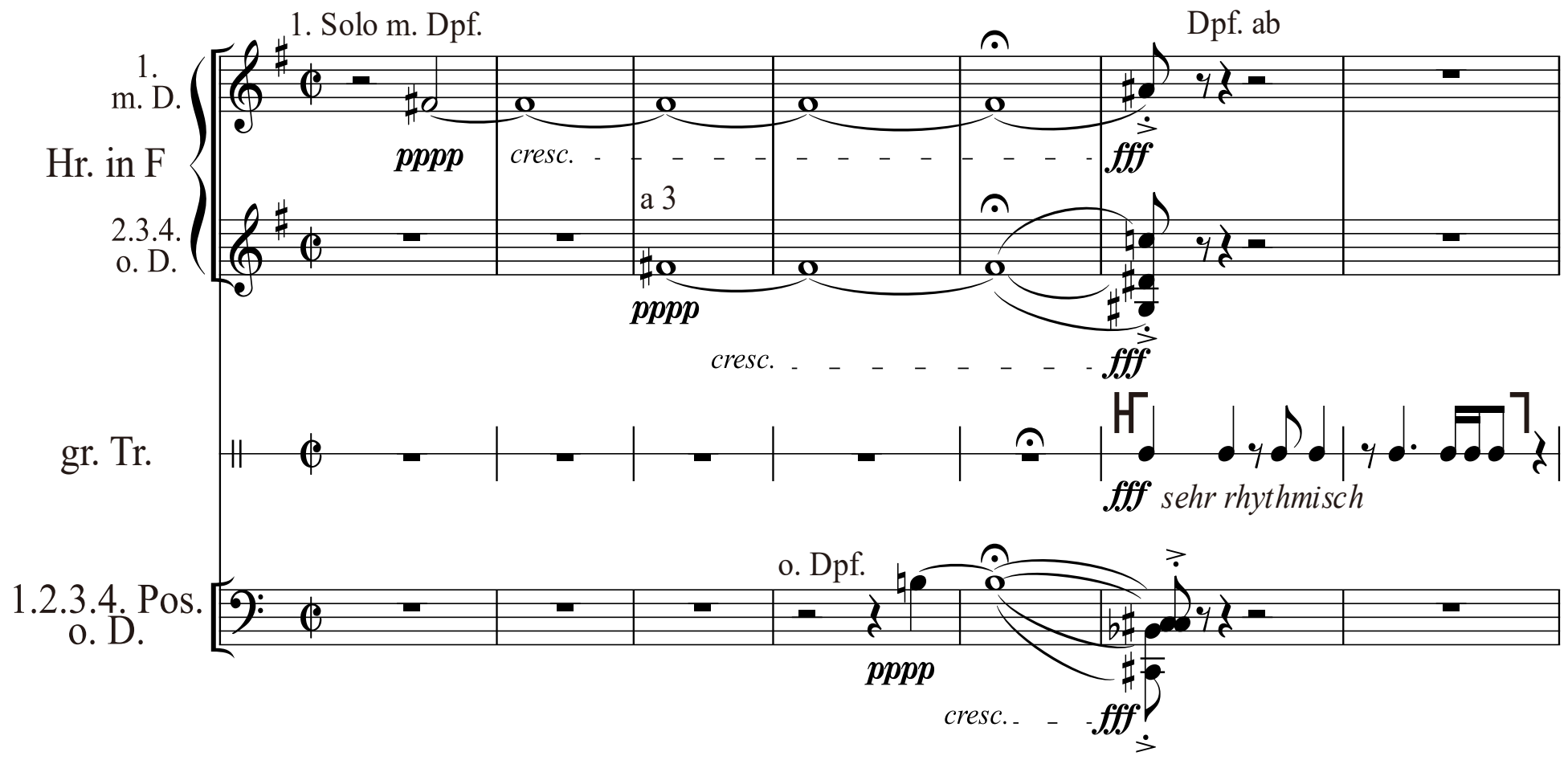

Fig. 8: A. Berg “Wozzeck” from Act 3 Scene 2
“Invention über einen Ton H”

## 3. ‘Klangfarbe Null Grades (The Zero degree of Timbre)’ … Instrumental accord from the Perspective of Gaussian Trial Timbre

In the previous section, we indicated the frequency of a sine wave—which produces the same pitch—as a reference for the pitch of instrumental sounds. W. Sethares also chooses the sine wave as a reference point, regarding it as the ‘origin’ of timbre.

If we assume that timbre is something intended to be heard by the human ear, then the ‘sine wave’ becomes something akin to absolute zero—an unattainable ideal. This is because a perfect sine wave cannot perform any physical work [18]; consequently, it cannot cause our eardrums to vibrate, nor can our auditory system perceive it.

Therefore, at the center frequency $f_0$, human hearing can detect a normalised Gaussian spike with a narrow line width proportional to $f_0$:

$$\sigma_0 = \frac{2^{\frac{1}{12}}}{2\sqrt{2\ln 2}} f_0 \quad \cdots ② - \mathrm{a}$$

is defined as

À Vinko Globokar, le maestro

$$\pi(f) = \mathcal{N}(f_0|{\sigma_0}^2) = \frac{1}{\sqrt{2\pi}\sigma_0}\exp\left[-\frac{(f-f_0)^2}{2{\sigma_0}^2}\right] \cdots ② - b$$

Let us prepare an ultra-narrowband Gaussian noise $\pi(f)$ with pitches equivalent to that of each constituent notes of the chord.

In the following, we shall refer to $\pi(f)$ as the 'Fletcher Gaussian' or 'Fletcher elementary spike', in honour of Harvey Fletcher—the true discoverer who laid the foundations of auditory perception and (contrary to popular belief, which attributes this to Robert Millikan, Fletcher's mentor) actually conducted the experiments to determine the elementary spike. Hereafter, we shall regard the Fletcher Gaussian as a vertex, calculate the Wasserstein distance between them, and present the results of mapping them onto a complete normed space (a Banach space) identical to that of instrumental timbres.

First, for the C♯ and F♯ notes played on the trombone in Mahler's Symphony No. 3, as shown in Fig. 1, we present a side-by-side comparison of the sounds with and without a mute, alongside Fletcher's 'unit timbre'.

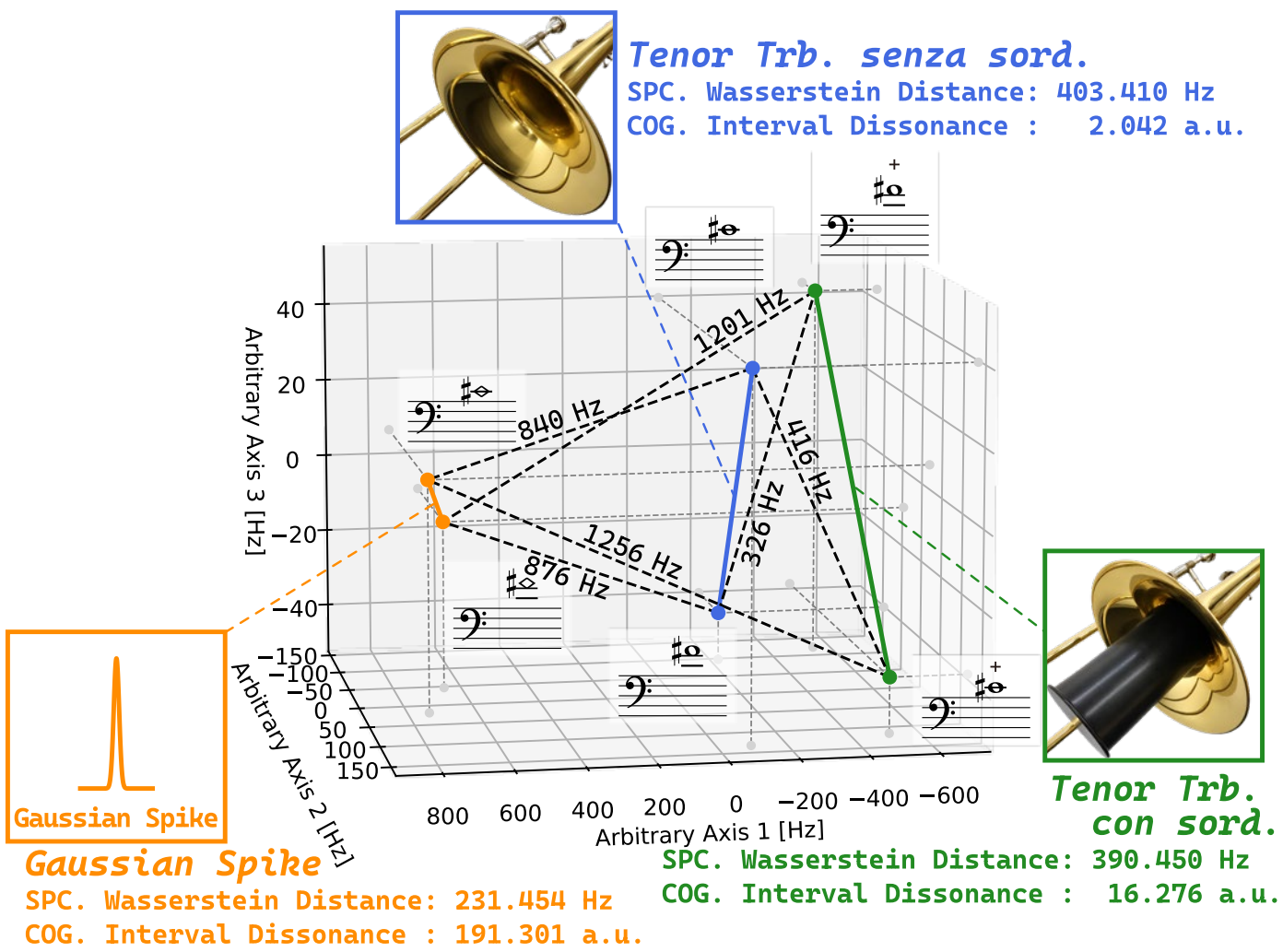

Fig. 1-c: Topological mapping of spectra: G. Mahler Symphony 3, 1 mvt. Bar. 341～ :
" C# - F# " Con-sord Simplex, Senza-sord Simplex & Gaussian Simplex

It should be noted that Fletcher noise approaches a sine wave as a limiting case, and that the Wasserstein distance between the corresponding vertices is therefore almost identical to the difference between their centre frequencies.

Similarly, in order to examine the relationship among three notes rather than between two, Fig. 2 shows the results of mapping the timbral fragments derived from Fletcher noise into a Banach space. The analysis is applied to the B♭-major chord in Mahler's *Symphony No. 3* shown in the figure, comparing the trombone sounds produced with and without mutes.

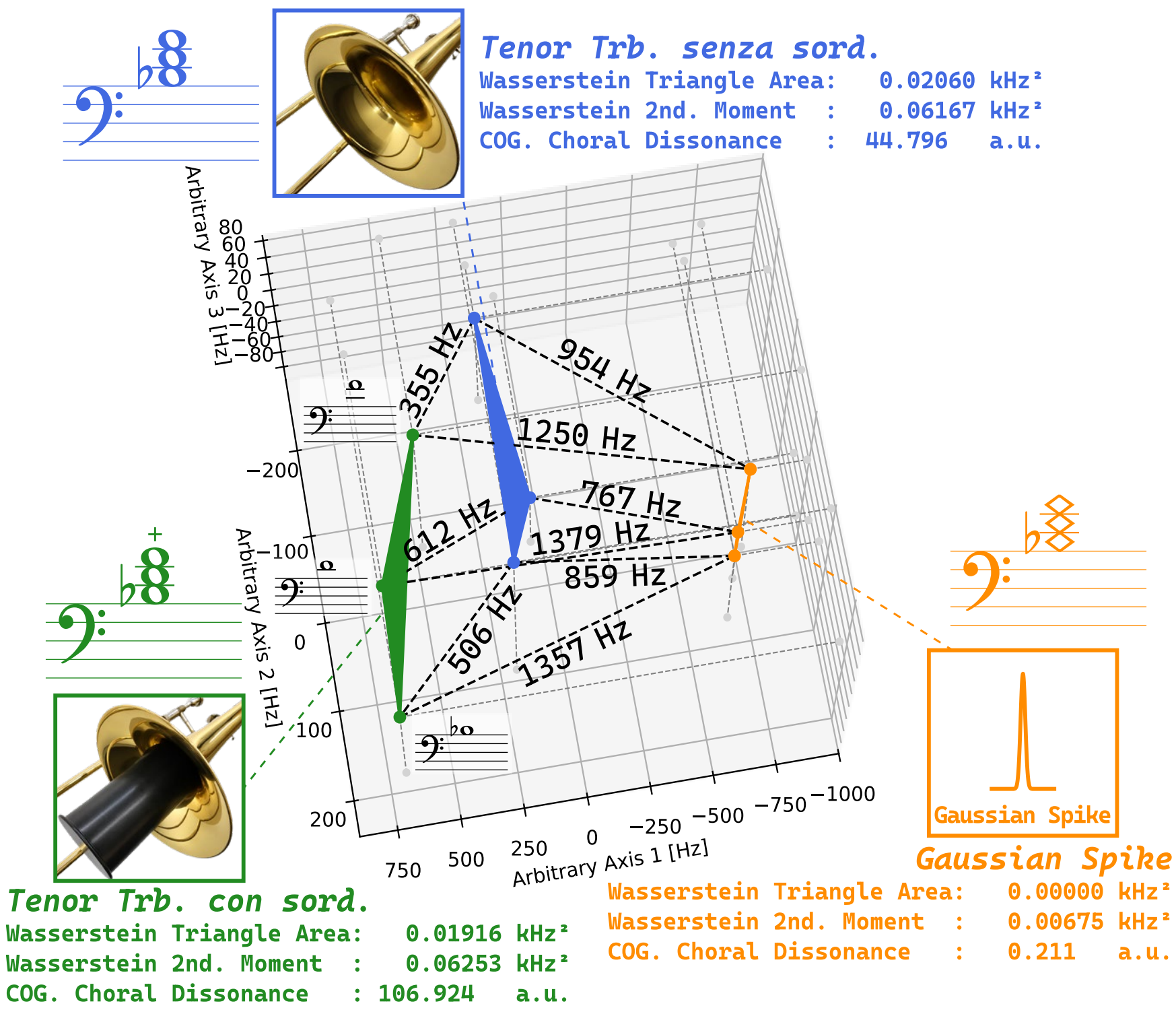

Fig. 2-c: Topological mapping of spectra : G. Mahler Symphony 3, 1 mvt. Bar. 341～ :
" B ♭ - D - F" Con-sord Complex, Senza-sord Complex & Gaussian Complex

The 'triangle' of the Fletcher-Gaussian spike, shown in orange, is extremely flat and has virtually no area. If it were a delta-function-type sinusoidal spike, the Wasserstein distance would be equal to the frequency difference; consequently, the Wasserstein triangle would take the form of a line segment divided internally, and its area would be zero. As the calculation above uses Fletcher noise of finite width, a finite value is obtained [19]; however, as this is aurally close to the threshold of perception, it would be appropriate to regard its area as effectively zero.

Traditional harmony, originating with J. Ph. Rameau, is a system developed on a grand staff that notates pitch alone. Consequently, these can be represented by sine waves or narrow-band noise. As these form (approximate) line segments in a Banach space, the line segment obtained in the limit of sine waves can be regarded as a common 'axis', and it becomes apparent that any timbre can be quantitatively evaluated in a probabilistic metric space as a deviation from the 'zero-timbre axis'. As will be confirmed immediately in the next section, these function as a common 'zero-timbre axis' leading to a more diverse range of timbres.

## 4. The Serial Integration of Klangfarbenmelodie and Sprechgesang

The years 1924–25, when Berg's *Wozzeck* was first performed, coincided with the introduction of the Harmon mute in Chicago, Illinois, USA.

This period also coincided with the premiere of George Gershwin's *Rhapsody in Blue* (1924), performed by Paul Whiteman's jazz band. Ferdie Grofé, who arranged the work, simply marked 'Wha-Wha mute' in the trumpet part, leaving the precise tonal variations to the performers' interpretation.

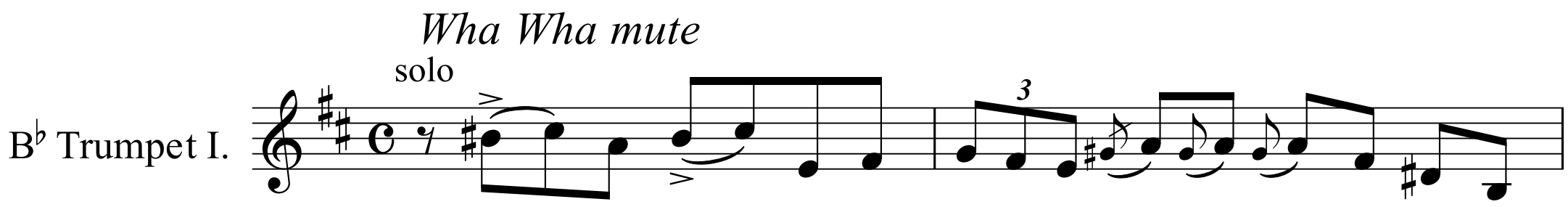

Fig. 9: Early examples of the use of the wow-wow mute in orchestral music:
G.Gershwin- F. Grofé Rhapsody in blue（1924）

The 'Wha-Wha' mute used here was originally a simple cleaning device known as a 'plunger'. The 'Harmon' mute, however, soon became the standard. George Schlusselburg, a Chicago resident, applied for a patent for the Harmon mute in 1924. The patent was granted in 1925, after which the mute came into use and rapidly gained popularity.

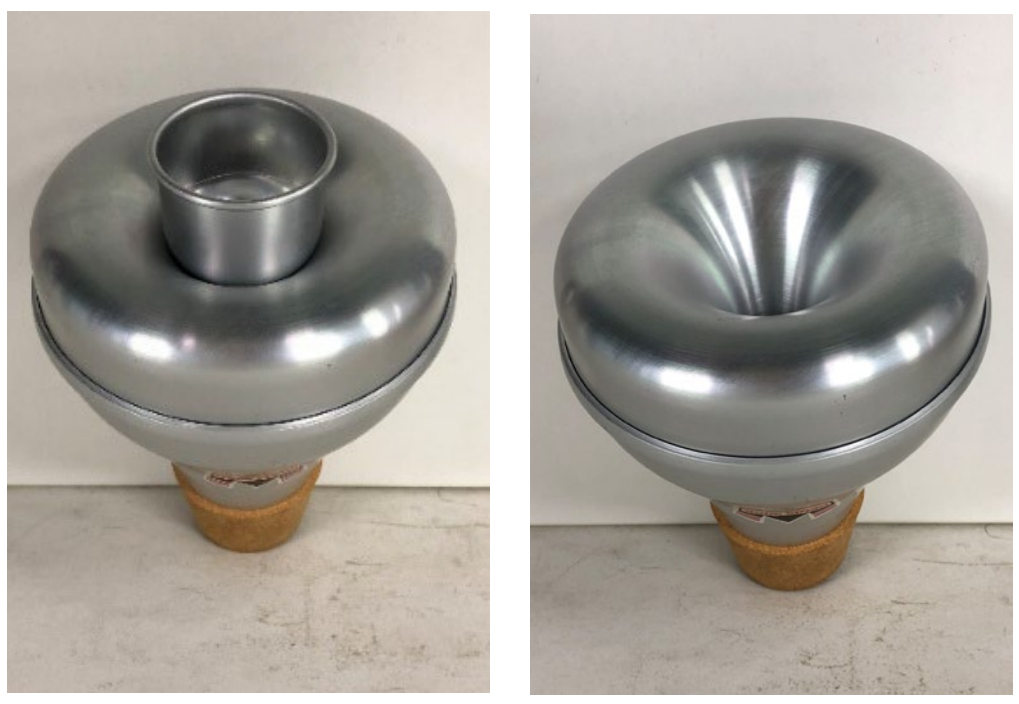

Fig.10: Harmon mute

This mute has a hole in the center, and the "tone color" is altered by opening and closing this aperture by hand. Whilst Harmon mutes are available for trumpets and trombones, and are also known for the French horn, they are not generally manufactured for euphoniums or tubas, as it is difficult to manipulate the bell by hand on these instruments. By opening and closing this aperture with the hand, the Harmon mute can produce a 'wow-wow' effect in the tone.

"Closing the opening of the Harmon mute completely" with the palm of the hand produces a timbral change similar to [u], whilst opening it produces a sound similar to [a]. Interestingly, if one shifts just a few finger(s) to create a gap—resulting in a 'half-open' position—the timbre changes to a sound resembling the vowel [o] in spoken language.

If we map the Wasserstein triangles onto a Banach space for the three-part trombone arrangement of Webern's Op. 1 chord shown in Fig. 4, played with the Harmon mute opening in the 'fully closed', 'half-open' and 'fully open' positions, the results are as follows.

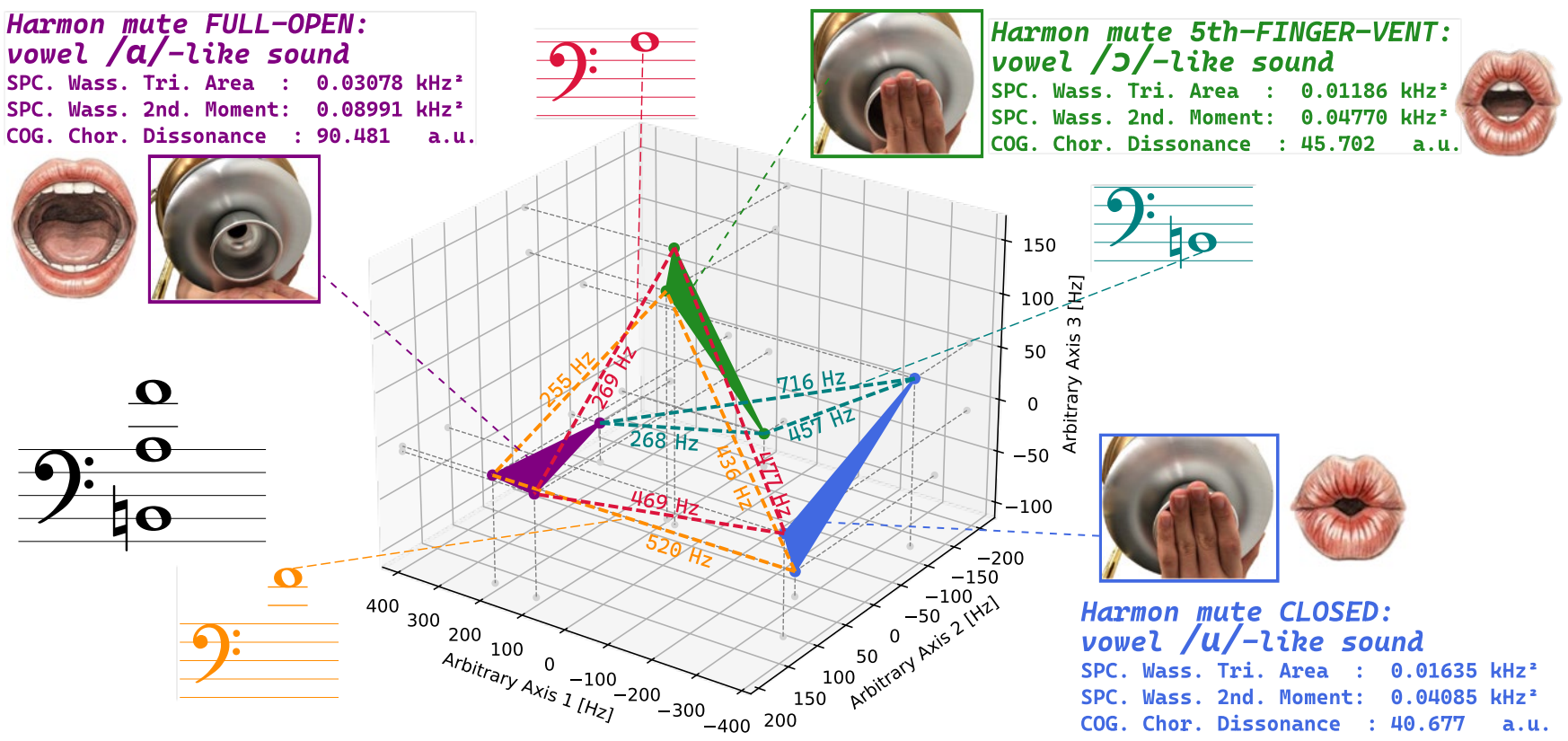

Fig. 4-b: Topological mapping of Syllabic spectra with Harmon mute:
A. Webern "Passacaglia" " H-A-F ": [A] Complex. [O] Complex & [U] Complex

Furthermore, if we map both the Fletcher noise triangle (which, due to its flattened shape, effectively becomes a line segment) and the intervals comprising similar pitches onto a Banach space, the result is as follows.

Just as Wagner was unable to use straight mutes for the trombone during his lifetime, the rise in pitch and the resulting instability in playability caused by the insertion of mutes were serious problems. Whilst custom-made products were used in Germany for R. Strauss's "Don Quixote", in the United States from the 1920s onwards, mutes for brass instruments saw rapid development and widespread adoption, originating in jazz bands.

When the Nazis seized power, Schönberg, who was Jewish, went into exile outside Austria, whilst Berg and Webern remained in Europe. Whilst mutes for string and brass instruments were heavily utilised in the works of G. Mahler and the Second Viennese School to achieve 'timbral changes', approaches such as the Harmon mute—which alter the mute itself—cannot be found in the works of this 'pioneering generation of timbral melody'.

Not only in the scores of Berg, who died prematurely in Austria in 1935, and Webern, who was accidentally shot and killed in 1945, but also in those of Schönberg—who lived in the United States until the 1950s—no specifications for mutes other than the orthodox straight mute can be found.

Now, let us reconsider the 'trombone sound resembling [a], [o] and [u]' from a different perspective as the 'pronunciation of the vowels [a], [o] and [u] by the trombone', and furthermore, let us map this, together with the Fletcher noise triangle (which, consisting of similar pitches, is flattened into what is effectively a line segment), onto a Banach space.

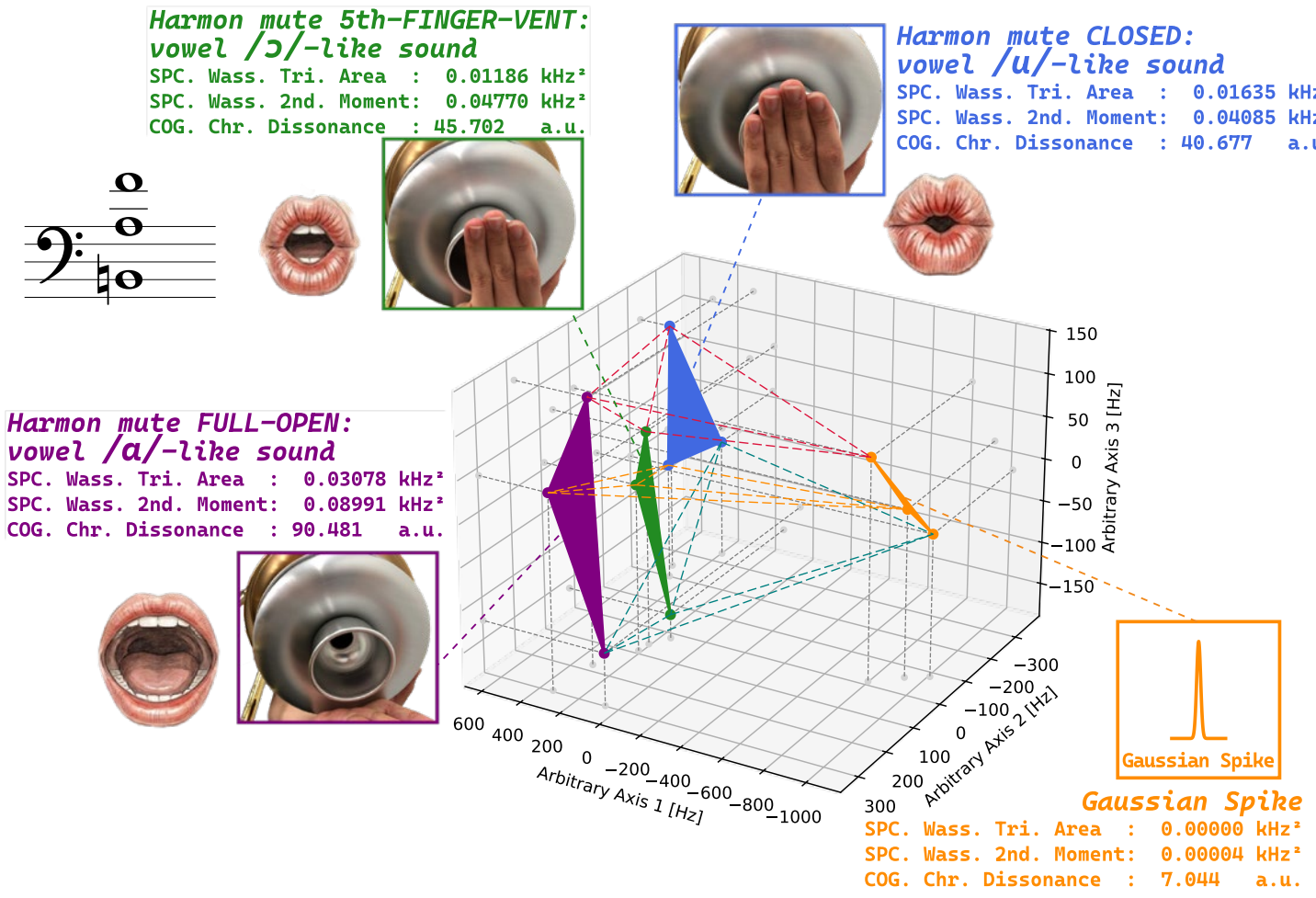

Fig. 4-b’: Topological mapping of Syllabic spectra with Harmon mut :
A. Webern “Passacaglia” “ H-A-F ”: [A] Complex. [O] Complex, [U] Complex & Gaussian Complex

It can be confirmed that the axis of the Fletcher noise—which might be termed the ‘zero degree of timbre’—is situated at a distance from each of the ‘triangle of timbre-chord’ formed by the trombone’s [a], [o] and [u] sounds.

From this, it can be confirmed that Fletcher-type ultra-narrow-band Gaussian noise—or, as its limit, a sine wave—is positioned not only as the ‘zero degree of timbre’ but also as the **‘zero of speech’** and the **‘zero degree of vowels’**.

Therefore, let us once again consider evaluating ‘human spoken language’ from this perspective using the same Wasserstein geometry.

We performed a topological mapping of the results of singing the middle A on the treble clef using the five vowels [a], [e], [i], [o] and [u], as well as humming [m], in a soprano voice. By selecting the display angle, it is possible to reproduce the well-known vowel triangle on a Banach space.

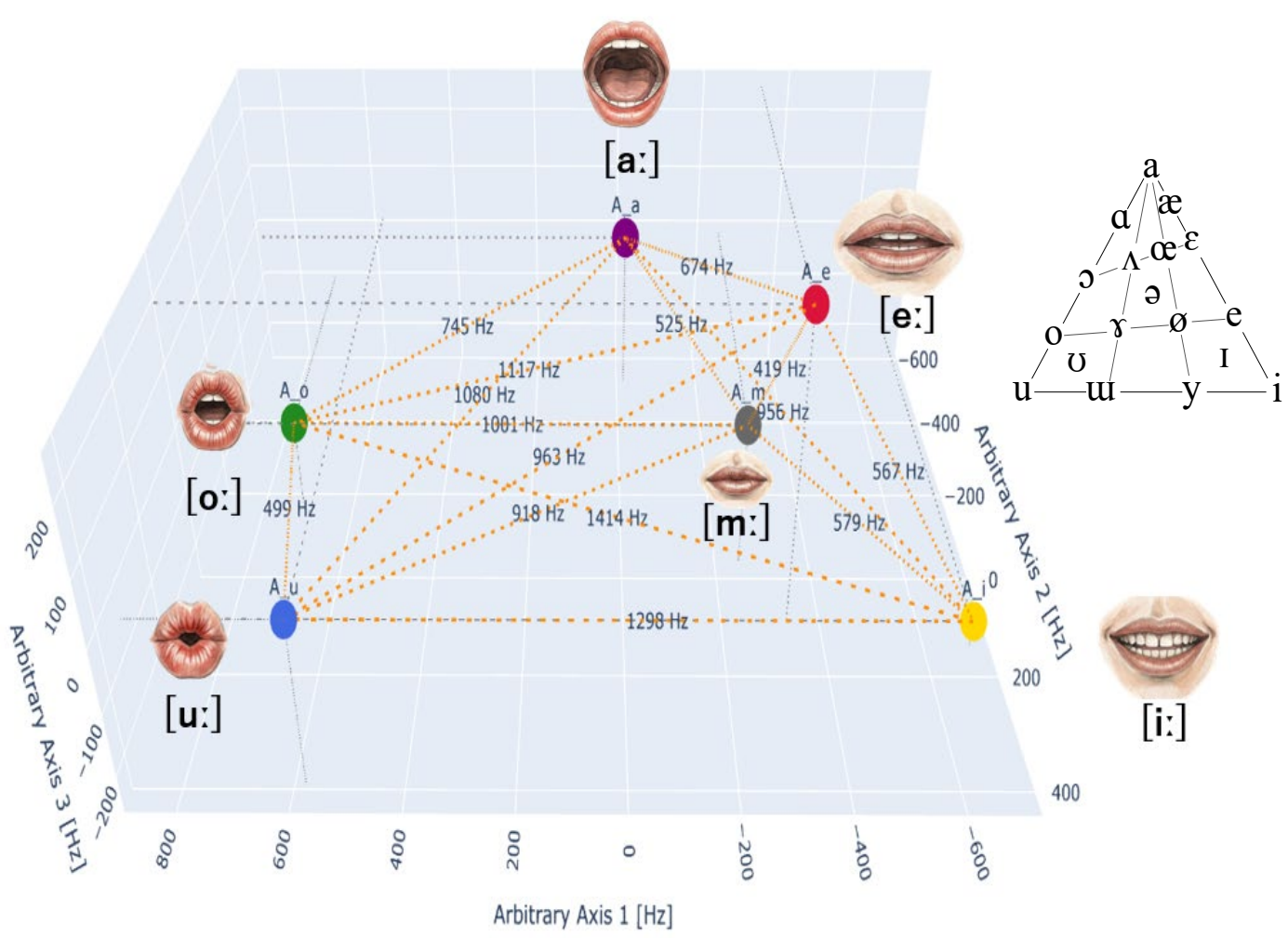

Fig. 11: Quantification of the vowel triangle

When this configuration is plotted on the triangle representing Webern's chord, shifted up by one octave, as shown in Fig. 4, the result is as follows.

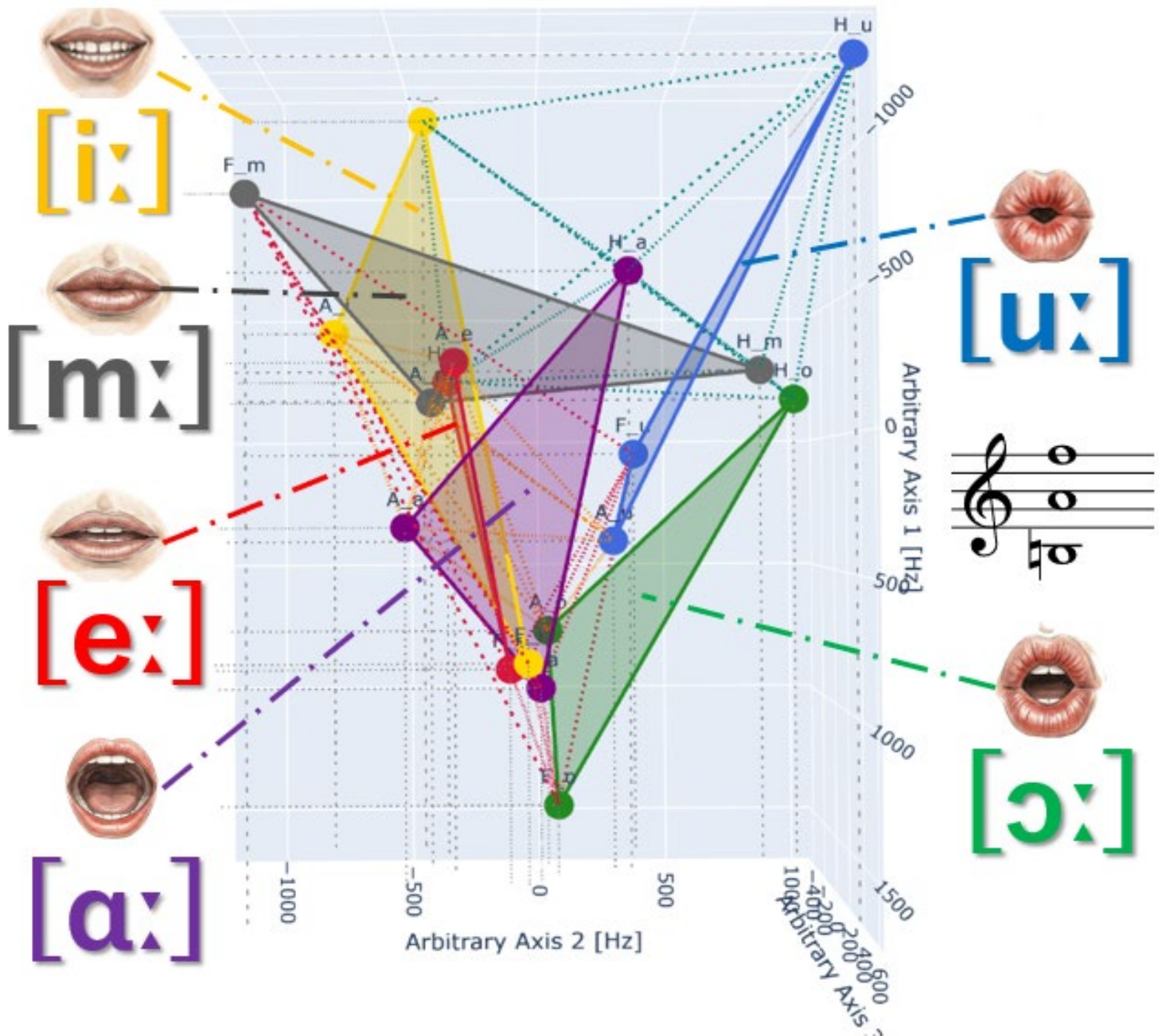

Fig. 4-c: Wasserstein-triangular representation of vowels with pitches.
on A, Webern "Passacaglia"'s " H-A-F "
"Gesang-klangfarben"

As these represent Wasserstein's geometric representation of 'vocal timbres accompanied by intervals', they can be interpreted—by extending Schönberg's concept—as a co-ordination of 'Gesang-Klangfarben' (interval-based vocal timbres).

A. Schönberg, who left Austria in 1933, took up a professorship at the University of California, Los Angeles, in 1936, took up residence in Beverly Hills, and became friends with his neighbor, G. Gershwin. The anecdotes of them enjoying tennis together are widely known. Gershwin died the following year, having completed and performed the majority of his major works; it is natural to assume that Schönberg had heard them. Put simply, it is highly likely that Schönberg himself was aware, in terms of sound, that the 'timbre'—or 'vowel'—changes produced by the opening and closing of the harmon mute, as commonly used in jazz, could be achieved on brass instruments. However, Schönberg himself did not think along these lines, and there is no evidence within his body of work that he viewed the opening and closing of the plunger or Harmon mute as relating to 'spoken language'. The same applies to Alban Berg and Anton Webern; indeed, the use of the wow-wow mute itself remained far from commonplace in orchestral music right up until the 21st century. The work of Ernst Ernst Krenek and Hans Eisler, amongst others, was premised on 'jazz', and it was not until the post-Second World War generation that an approach emerged which directly engaged with the changes in sound themselves.

So, has this Viennese approach to music come to an end? We would hesitate to say so. One witness to the contrary is Beat Furrer (1954-), the Swiss-born Austrian composer and conductor, who continues to work in Vienna today. His music is distinguished by a style that systematically brings noise into the fabric of musical sound. In the genealogy of musical thought, his work may be seen as a living inheritance: the youthful aspirations of Gustav Mahler and Arnold Schönberg, preserved and carried forward into the twenty-first century.

It is essential to acknowledge the immense contribution made by Vinko Globokar (1934-) to the exploration of the underlying relationship between musical timbre and spoken language [20]. During the early days of the French National Institute for Music and Acoustics (IRCAM), Pierre Boulez appointed Globokar as director of the 'Instrument and Speech Research Department', and practical investigations into the boundary between instrumental sound and spoken language were carried out in collaboration with István Matuz (flute) and Heinz Holliger (oboe), amongst others.

Boulez had in mind Schönberg's three problems, which were the subject of vigorous debate in the 1950s, particularly the 'Sprechgesang problem' and the 'Klangfarbenmelodie problem', and his approach focused on their common ground. However, due to the limitations of knowledge at the time, analogue sound processing, and computer technology in the 1970s, the discussion diverged from the behavior of sound itself towards physicality involving action, literary connotations, and even social and political expression through the 'voice', and thus failed to reach sufficient depth.

Here, in homage to Vinco Globokar,we shall plot the trombone's [a], [o] and [u]—played with the Harmon mute fully closed, half-open and fully open respectively—alongside the soprano's [a], [o] and [u], sung an octave higher on the same note names, within the same Banach space.

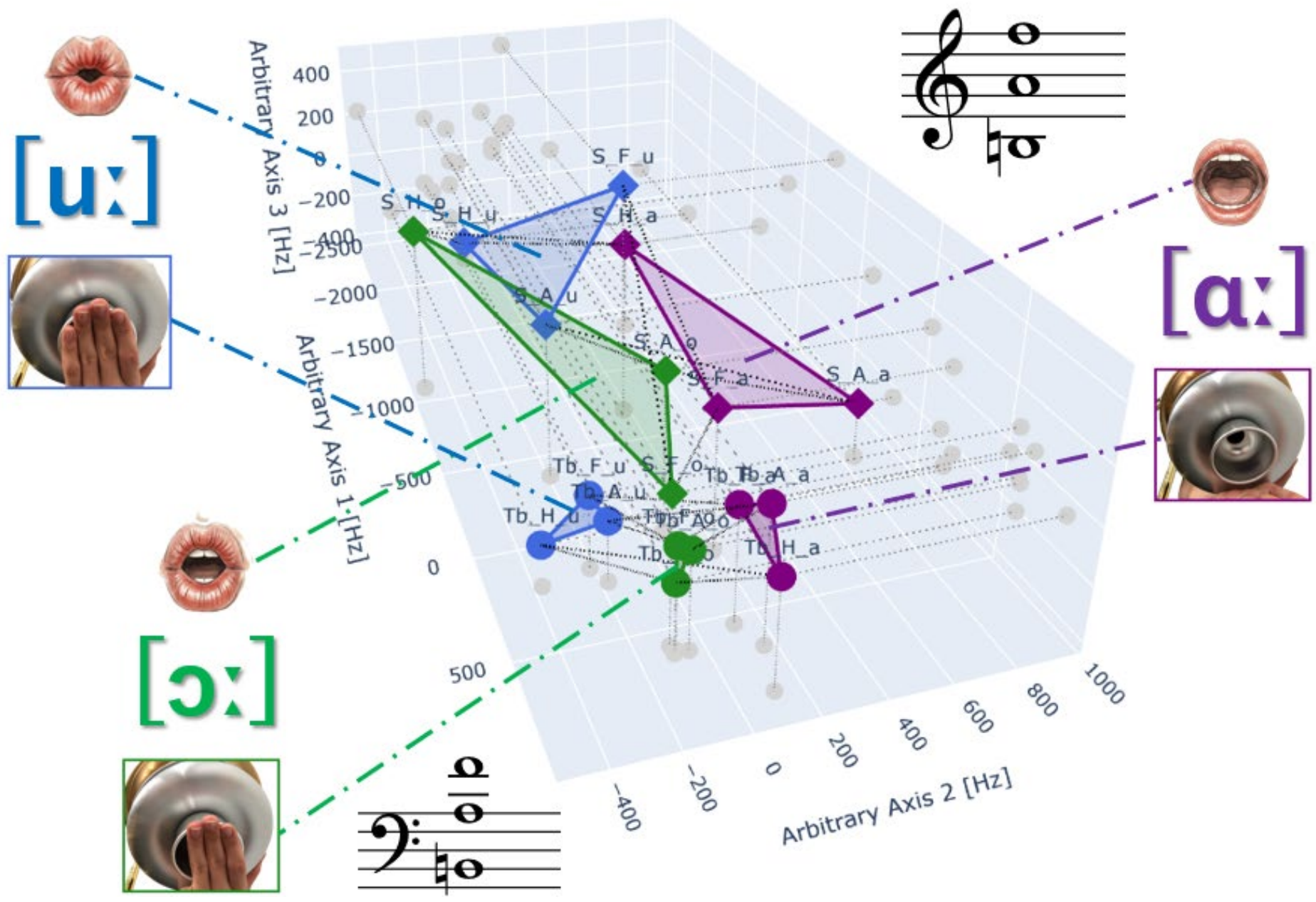

Fig. 12: Serial unification of Sprechgesang Klangfarbenmelodie
on A. Webern "Passacaglia"'s " H-A-F "

Although an analysis of ‘vowels’ produced by brass instruments and human speech sounds within the same acoustic space is beyond the scope of this article and will be addressed in a subsequent article, various correspondences can already be identified between syllables and pitch-defined polyphonic structures.

The digital processing technologies mentioned at the beginning of this article—including Kurzweil’s sampling synthesizer and automatic text-to-speech systems—came to the fore from the 1980s onwards, while the issues that had been explored before the 1970s largely fell outside the scope of these technologies [21].

The question raised in this paper is by no means a new one. Rather, it is a problem that has been posed repeatedly over the past two or three centuries, yet has never been adequately formulated.

Let us take a closer look at the subtitle of Helmholtz’s seminal work, published in 1863: *Die Lehre von den Tonempfindungen*, or *On the Sensations of Tone*, whose subtitle reads *als physiologische Grundlage für die Theorie der Musik* (‘as a Physiological Basis for the Theory of Music’).

Neither Lord Rayleigh (John William Strutt), Chandrasekhara Venkata Raman, nor György von Békésy showed a sustained interest in elucidating the sense of tonal progression in just intonation—a subject to which Helmholtz, a pianist with absolute pitch, devoted considerable attention. This, however, in no way diminishes the value of these unresolved questions. Rather, it points towards a vast and largely uncharted field in which ‘the theory of music’, in its fullest sense, remains to a remarkable extent unexplored.

## 5. Brief Concluding remarks, Summary of Unresolved Issues and Acknowledgements

The progress of this thesis corresponds closely to the point reached by Helmholtz in the middle of the 1850s. In establishing physiological acoustics, Helmholtz developed a method for evaluating dissonance through the phenomenon of ‘beating’. By around 1859–61, he had largely completed his analysis of speech sounds and his synthesis of vowels using tuning forks. However, a fully functional version of his vowel-tuning-fork system, which employed electromagnets, was not achieved until 1905, after his death.

With regard to instrumental timbre, the present research has reached the stage at which the fundamental framework for representing spectra, quantified in hertz, by means of Wasserstein geometry has been established. A detailed examination of the specific treatment of timbre will be presented in a subsequent paper. A separate paper is also being prepared on the theoretical foundations and mathematical formulation of the method.

With regard to speech sounds, the present study has so far been limited to the classification of a restricted set of vowels. The treatment of consonants, attack characteristics, and other aspects of speech will require further investigation.

These issues relate to the three problems posed by Schönberg at Luigi Nono's seminar in 1987, namely:

1. **The treatment of sound—the serial approach**
2. **The distinction between singing and speaking**
3. **Klangfarbenmelodie**

After approximately forty years of engagement with these three problem formulations, the present paper demonstrates, in a preliminary manner, that Problems 2 and 3 are essentially manifestations of the same underlying question, and that their treatment can be integrated with Problem 1. More specifically, the serial procedure envisaged here is a sequential mapping of spectral vertices based on the Wasserstein distances between spectra. In this sense, the present study suggests that a unified serial treatment of timbre, speech, and melody is mathematically feasible.

We should also note that the Wasserstein distance is defined here not in physical real space, but in the probability space of a normalised spectrum. In other words, although we perceive music and speech through our ears as physical phenomena occurring in real space, the present approach can become fully effective only when the problems of spatial acoustic perception—including both auditory and cerebral processes—are properly taken into account. Only then can this framework contribute to an understanding of the 'wisdom of music' as it actually resounds in the physical world.

Finally, we would like to express our deepest gratitude to Mme Masako Shindo (soprano), Ms Hinako Tani (horn), Mr Taishi Yoshizawa (trombone), and Mr Keito Teshima (bass trombone) for their invaluable cooperation in recording the sound sources used in this research. We would also like to express my profound gratitude to the late Jōji Yuasa, the late Toshirō Mayuzumi, the late Luigi Nono, the late Pierre Boulez, the late Karlheinz Stockhausen, the late Jean-Claude Éloy, the late Peter Eötvös, the late John Pierce, the late Max Mathews, and the late Hiroshi Inose, whose work and thought helped to identify and illuminate the problems addressed in this paper. We are equally grateful to Jin-young Lee and Pengyue Yuan, with whom we had valuable discussions during the preparation of this manuscript, as well as to Vinko Globokar, Professor Minoru Tsukada, and Professor Shun'ichi Amari.

□

# Notes and references

[3] Ray Kurzweil. “The Singularity Is Near: When Humans Transcend Biology”. Viking Penguin (2005).

[4] In the early 1980s, Japan and the United States collaborated to establish the MIDI standard; in France, a team at IRCAM at the Centre Pompidou, led by P. Boulez, was developing the 4X hardware system; and Kurzweil also developed a system using his own hardware that allowed users (in this case, Stevie Wonder) to record the sounds of instruments in front of them and play them back on an electronic keyboard.

[5] On 30 November 1987, at a lecture by Luigi Nono (1924–90) held at the Faculty of Liberal Arts, University of Tokyo. This lecture was organized by one of the authors whilst he was a student. Nono harshly criticized the development of computer music from the 1970s onwards as ‘superficial endeavors geared towards commercialization’. Arnold Schönberg, who was Nono’s father-in-law, believed that pitch was merely one aspect of timbre, and that timbre itself was essential. He argued that the choice of timbre in instrumental composition should not be viewed merely as an effect, but should be understood more structurally and serve the musical structure.

Consequently, in the 1950s, an approach to timbre directly linked to ‘musical structure’ was vigorously explored, with Nono, Boulez, Stockhausen and others attempting various models. However, in the 1950s—when only analogue audio processing existed—the outcome was limited to the concept of ‘formants’, a term derived from speech. Consequently, Nono’s main criticism is that, once digital processing became possible in the 1970s onwards, interest in Schönberg’s questions—which at first glance appear outdated—dwindled, and fundamental research that fell outside the scope of commercial applications ceased to be pursued.

[6] Natalie Bauer-Lechner. „Erinnerungen an Gustav Mahler“. Herausgegeben von J. Killian. *E. P. Tal & Co. Verlag* (1923).

[7] When considering mutes for brass instruments in the 1890s—particularly the trombone—it is necessary to take historical circumstances into account. Throughout the lifetime of R. Wagner (1813–84), trombone mutes were made of wood or leather; when inserted into the bell, they tended to raise the pitch by approximately a semitone to a whole tone, resulting in a loss of control over both timbre and pitch. It is likely against this backdrop that almost no instances of trombone mutes can be found amongst Wagner’s vast body of scores.

The horn stands in stark contrast to this. A technique for manipulating pitch by inserting the hand into the bell had long been developed for the horn, and a tradition of producing a variety of pitches and timbres on the natural horn—as in Mozart’s Horn Concerto—had been firmly established.

In contrast, the trombone—which was designed to accommodate a wide range of key changes via its slide (particularly in German Protestant churches to reinforce the male voices in the choir)—offered no scope for the development of such techniques.

[8] “Don Quixote” was premiered in Cologne—where Franz Wüllner (1832–1902) served as director

of the conservatoire—under his baton; the composer was not present. However, it is thought that Richard Strauss had commissioned a set of mutes for brass instruments—which were rare at the time—and that these may have been loaned from Munich. It is also worth noting that Mahler's Symphony No. 3 was premiered in 1902 by members of the same Gürzenich Orchestra in Cologne. R. Strauss was not only well-versed in the various instruments himself, but was also closely connected to a network of brass players through his father, Franz Strauss (1822–1905), who was principal horn player with the Munich Court Opera Orchestra. Whilst anecdotes circulate about the elder Strauss's feud with R. Wagner, it is said that without Franz Strauss's contributions—notably the 'Call of Siegfried'—Wagner's operas would not have taken their present form. It is said that Franz Strauss gave his son Richard (who, by a twist of fate, came to share the same name as Wagner) a rigorous classical musical education modelled on Mozart. This also implied a thorough understanding of the natural horn playing technique prevalent in Mozart's time; the solo part of the Horn Concerto No. 1, composed for his father at the age of eighteen, displays a masterful style that demonstrates a command of both the natural horn and the valve horn, despite being the work of a teenager.

[9] R. Strauss, a conductor and composer well-versed in the characteristics of brass instrument mouthpieces, demonstrated great skill in his role as Kapellmeister at the Munich Court Opera—where his father had served as principal horn player for many years—by commissioning improvements to instruments and mutes, as well as the production of bespoke items. It may have been ironic for Franz Strauss that R. Strauss came to contribute to developments in orchestration—which Wagner himself had been unable to achieve—through the bell technique of the horn, a technique with which Wagner was not necessarily familiar. However, the result—the truly earth-shattering orchestration realised from *Don Quixote* onwards—must be regarded as a major achievement in the history of music.

[10] Although this is often misunderstood, the purest form of 'Klangfarbenmelodie'—as defined—is an approach that seeks to achieve musical expression by varying the timbre whilst keeping the same pitch (the height of the sound) constant. In the musical example by Mahler shown in Fig. 7, a 'timbre change' is notated in which the timbre of the horn played using standard technique decresces, whilst the sound of the same pitch produced by the horn using the stop technique crescendes. There were few deliberate examples of such musical ideas prior to Wagner; the concept was developed in the process of refining passages that had been prophetically notated through performances of works such as *The Ring of the Nibelung*.

The same development applies to 'Sprechgesang' (spoken song): through character-based singing (for example, the dwarf Mime in "The Ring of the Nibelung"), a dramatic form of 'spoken' expression—in which it was acceptable to sing off-pitch—became widespread, leading from there to the 'Verismo' (a vocal technique of 'speaking-singing' (Sprechstimme) that spread rapidly following P. Mascagni's "Cavalleria rusticana" (1890), based on a play by G. Verga) became a subject of debate, and exploration began.

[11] For details of the calculations, see [2].

[12] In Fig. 1-b, the physical entropy is defined as the Shannon entropy of a discrete spectrum:

$$S[p] = -\sum_i p(f_i) \log_2 p(f_i) \ \cdots ③ - \mathrm{a}$$

In the case of a sine wave where $p(f_i) = \delta(f_i - f_0)$, $p(f_i)$ can only take values 0 or 1. Consequently, the physical entropy is

$$S[\delta(f_i - f_0)] = 0 \ [\mathrm{bit}] \ \cdots ③ - \mathrm{b}$$

In actual data calculations, because the sample length $T$ of the sound source is finite, the discrete spectrum of a sine wave does not take the form $\delta(f_i - f_0)$ when $f_0 T$ is not an integer; however, with hypothetical infinite-length data, as mentioned above, a finite, determinate entropy value does not arise.

[13] Sethares, W. A. Local consonance and the relationship between timbre and scale. *The Journal of the Acoustical Society of America*, **94**(3), 1218–1228 (1993).

[14] The quantities of cognitive dissonance calculated in this paper follow [13]. In Sethares's model, the spectral peaks extracted from each sound document are referred to as *partials* and denoted by

$$i \equiv (f_i,\ a_i) \ \cdots ④ - \mathrm{a}$$

where $f_i$ denotes the frequency of a partial, while $a_i$ denotes its relative amplitude, with the maximum peak normalized to 1.

The Sethares dissonance between two partials $(f_i, a_i)$ and $(f_j, a_j)$ is defined as

$$d_{ij} \equiv \min(a_i, a_j) \left[ \mathrm{e}^{-B_1 s_{ij} |f_i - f_j|} - \mathrm{e}^{-B_2 s_{ij} |f_i - f_j|} \right] \ \cdots ④ - \mathrm{b}$$

where

$$s_{ij} \equiv \frac{S^*}{S_1 \min(f_i, f_j) + S_2} \ \cdots ④ - \mathrm{b}'$$

and $B_1 = 3.5$, $B_2 = 5.75$, $S^* = 0.24$, $S_1 = 0.0207$, $S_2 = 18.96$ Hz, respectively.

The sum of this $d_{ij}$ is taken, and the resulting value is calculated as the cognitive dissonance according to the number of voices.

In Fig. 1-b, the quantity $D_A$ calculated for a single tone as follows is shown in the figure as *cognitive vertex dissonance*.

Let the set of all partials of a given timbre spectrum $A$ be

$$\mathcal{P}_A \equiv \{ i = (f_i, a_i) \} \ \cdots ④ - \mathrm{c}$$

where, for each pair of partials $i, j$ ($i \neq j$, $i < j$, without loss of generality), taking the sum of all Sethares dissonances $d_{ij}$ calculated for these pairs gives

$$D_A \equiv \sum_{\substack{i,j \in \mathcal{P}_A \\ i<j}} d_{ij}, \ \cdots ④ - \mathrm{c}'$$

which is used to evaluate the cognitive dissonance of a single tone $A$.

In Figs. 1-b’, 7-b, 7-c, 7-d, 7-e, and 2-c, the quantity $D_{AB}$ calculated for two tones of the same timbre at a non-unison pitch interval is shown as *cognitive interval dissonance*.

Let the sets of all partials of two given timbre spectra $A, B$ be

$$\mathcal{P}_A,\ \mathcal{P}_B \quad \cdots ④ - \mathrm{d}$$

respectively. And, for their summed spectrum $\mathcal{P}_A \cup \mathcal{P}_B$, take the sum of $d_{ij}$ in the same manner as above. Here, for identical frequencies $f_i$, the amplitudes of the two spectra are simply summed.

$$D_{AB} \equiv \sum_{\substack{i,j \in \mathcal{P}_A \cup \mathcal{P}_B \\ i<j}} d_{ij} \quad \cdots ④ - \mathrm{d}'$$

This $D_{AB}$ serves as an index of the dissonance associated with the interval between two tones of the same timbre.

Furthermore, in Figs. 2-c, 4-b, and 4-b’, for three-voice chords consisting of tones of the same timbre, the quantity $D_{ABC}$ calculated as follows is shown as *cognitive choral dissonance*.

Let the sets of all partials of the three timbre spectra be

$$\mathcal{P}_A,\ \mathcal{P}_B,\ \mathcal{P}_C \quad \cdots ④ - \mathrm{e}$$

and take the sum of $d_{ij}$ in the same manner as above.

$$D_{ABC} \equiv \sum_{\substack{i,j \in \mathcal{P}_A \cup \mathcal{P}_B \cup \mathcal{P}_C \\ i<j}} d_{ij} \quad \cdots ④ - \mathrm{e}'$$

This $D_{ABC}$ serves as an index of the dissonance associated with the intervals between three tones of the same timbre.

In all cases, the same operation is performed as in calculating the *cognitive vertex dissonance* for a single spectrum obtained by adding the spectra of all consisting source tones.

[15] The Wasserstein distance satisfies the triangle inequality.

$$W(P,R) \leq W(P,Q) + W(Q,R) \quad \cdots ⑤$$

Therefore, for a three-voice chord, a Wasserstein triangle can always be defined.

However, for four or more voices, it is not always possible to construct a consistent “Wasserstein tetrahedron” whose six edges are given by the six Wasserstein distances. We give one counterexample below.

Using the Dirac delta function $\delta(x)$, consider the four probability density functions

$$P(x) = \frac{1}{2}\delta(x) + \frac{1}{2}\delta(x-2),$$

$$Q(x) = \delta(x-1),$$

$$R(x) = \delta(x-2),$$

$$S(x) = \frac{1}{2}\delta(x-1) + \frac{1}{2}\delta(x-2).$$

The Wasserstein distances between these distributions are $\overline{PQ} = \overline{QR} = \overline{RP} = 1$, $\overline{PS} = \overline{QS} = \overline{RS} = \frac{1}{2}$. As is evident from the figure, these six edge lengths cannot form a consistent tetrahedron.

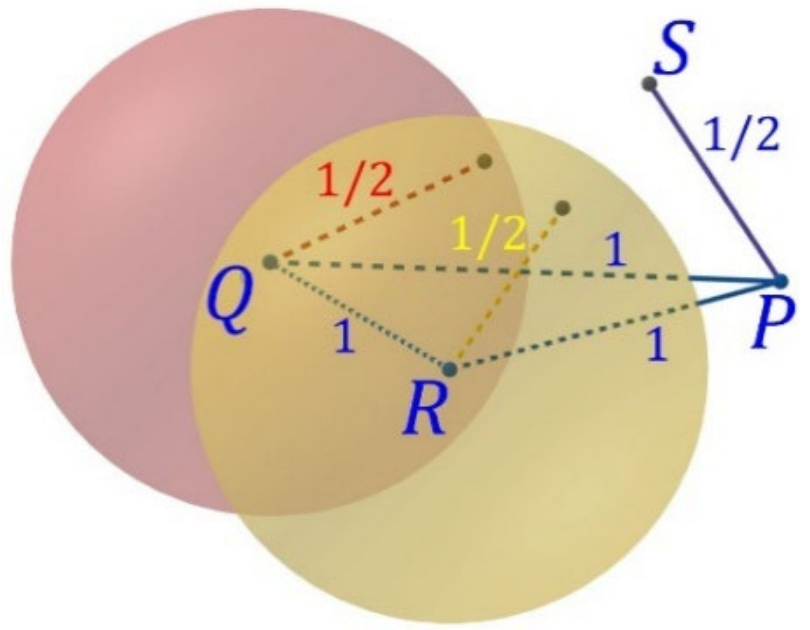

Fig. 13: Counterexample to consistent tetrahedron construction via Wasserstein distances of four probability distributions $P$, $Q$, $R$ and $S$.

[16] In preparation for evaluating the timbre of chords consisting of four or more notes (see [15]), we forumlate the Wasserstein second moment for three-voice chords as follows. Let the three timbre probability distributions be $p_1$, $p_2$ and $p_3$, and let their "Wasserstein center of mass" be $\tilde{p}$, defined as

$$\frac{1}{3}\sum_{1\le i<j\le 3} W_1\left(p_i, p_j\right)^2 = \sum_{i=1}^{3} W_1(p_i, \tilde{p})^2. \quad \cdots ⑥ - \mathrm{a}$$

The left-hand side defines the Wasserstein second moment,

$$I(\{p_1, p_2, p_3\}) = \frac{W_1(p_1, p_2)^2 + W_1(p_2, p_3)^2 + W_1(p_3, p_1)^2}{3}, \quad \cdots ⑥ - \mathrm{b}$$

and represents the mean squared pairwise Wasserstein distance among the three timbres. The right-hand side

$$W_1(p_1, \tilde{p})^2 + W_1(p_2, \tilde{p})^2 + W_1(p_3, \tilde{p})^2, \quad \cdots ⑥ - \mathrm{c}$$

measures how widely the three timbres are dispersed around their "central timbre," $\tilde{p}$.

In other words, the equality of these two quantities shows that the internal variation of a set of timbres $\{p_1, p_2, p_3\}$, expressed as the sum of squared distances from its Wasserstein center of mass, can be obtained solely from the edge lengths of the simplex, without explicitly calculating the Wasserstein center of mass $\tilde{p}$.

Furthermore, we can extend the definition of the Wasserstein second moment to general $n$-voiced chords as follows.

$$I(\{p_k\}_{k=1}^{n}) = \frac{1}{\binom{n}{2}} \sum_{1\le i<j\le n} W_1\left(p_i, p_j\right)^2, \quad \cdots ⑥ - \mathrm{d}$$

where $\binom{n}{2}$ is the number of pairwise distances.

[17] The area and second moment of a Wasserstein triangle exhibit a strong positive correlation, as shown below.

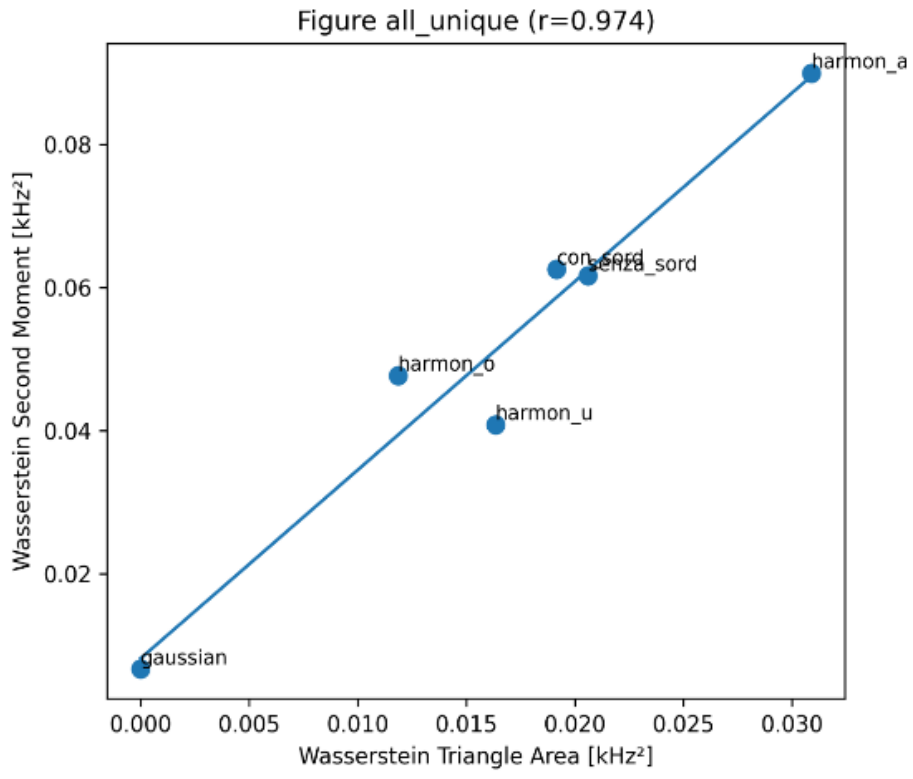

Fig. 14 Correlation between the area of Wasserstein triangles and the square choral moment

In this paper, we examine Wasserstein triangle areas for geometrical intuition. Note that, as shown above, similar considerations apply to more complex polyphonic structures involving four or more voices.

[18] Consider a real signal $g(t)$ (with finite energy: $\int |g(t)|^2\,\mathrm{d}t < \infty$) and its Fourier transform $G(\omega) = \int g(t)\mathrm{e}^{-\mathrm{i}\omega t}\,\mathrm{d}t$. The temporal and angular-frequency spreads—i.e. the fluctuations—are given by

$$\sigma_t^2 = \frac{\int t^2 |g(t)|^2\,\mathrm{d}t}{\int |g(t)|^2\,\mathrm{d}t} \quad \cdots ⑦-\mathrm{a}, \quad \sigma_\omega^2 = \frac{\int \omega^2 |G(\omega)|^2\,\mathrm{d}\omega}{\int |G(\omega)|^2\,\mathrm{d}\omega} \quad \cdots ⑦-\mathrm{b}.$$

Then, by Parseval's theorem

$$\int (g')^2\,\mathrm{d}t = \frac{1}{2\pi}\int \omega^2 |G|^2\,\mathrm{d}\omega \quad \cdots ⑦-\mathrm{c}.$$

Here, integrating by parts, we obtain

$$\int t g(t) g'(t)\,\mathrm{d}t = -\frac{1}{2}\int g(t)^2\,\mathrm{d}t \quad \cdots ⑦-\mathrm{d};$$

therefore, applying the Cauchy–Schwarz inequality, we obtain

$$\left(\int (tg)(g')\,\mathrm{d}t\right)^2 \le \int (tg)^2\,\mathrm{d}t \int (g')^2\,\mathrm{d}t \quad \cdots ⑦-\mathrm{e},$$

hence,

$$\frac{1}{4}\|g\|^4 \le \sigma_t^2 \|g\|^2 \cdot \sigma_\omega^2 \|g\|^2 \quad \Longrightarrow \quad \sigma_t \sigma_\omega \ge \frac{1}{2} \quad \cdots ⑦-\mathrm{f},$$

thus, we obtain Weyl's inequality, which gives a lower bound for the product of the fluctuations $\sigma_t$, $\sigma_\omega$ corresponding to zero-point fluctuations. This reveals a mathematical structure equivalent to that of the uncertainty principle in quantum mechanics.

If we consider the limit of a time fluctuation $\sigma_t \to \infty$ (a pure sine wave extending eternally, with neither a beginning nor an end), then $\sigma_\omega \to 0$, and the spectrum converges to $G(\omega) = 2\pi\delta(\omega - \omega_0)$, becoming a delta function with zero bandwidth. In this limit, the signal does not possess finite energy and therefore does no work. Consequently, a truly pure sine wave is a hypothetical vibration that can never be heard, since it does not cause the eardrum of a living organism to vibrate. In reality, sounds perceived by humans always begin and end at finite times (and thus possess a finite and definite value of fluctuation, $\sigma_t$). Consequently, $\sigma_\omega \neq 0\ (> 0)$, and a perceived sound is not a single-frequency component but a superposition of frequency components with finite linewidths. It is therefore necessary to consider a "wave packet" accompanied by fluctuations.

[19] The area $A$ of the Wasserstein triangle formed by three normalized Gaussian distributions $\mathcal{N}(\mu_1|\sigma_1^2)$, $\mathcal{N}(\mu_2|\sigma_2^2)$, $\mathcal{N}(\mu_3|\sigma_3^2)$, is given by Heron's formula as

$$A = \frac{1}{4}\sqrt{(D_{12} + D_{23} + D_{31})(-D_{12} + D_{23} + D_{31})(D_{12} - D_{23} + D_{31})(D_{12} + D_{23} - D_{31})} \quad \cdots ⑧ - \mathrm{a}$$

where

$$D_{ij} = W_1\left(\mathcal{N}(\mu_i|\sigma_i^2), \mathcal{N}(\mu_j|\sigma_j^2)\right) = |\sigma_i - \sigma_j| \left[\sqrt{\frac{2}{\pi}}\, e^{-\frac{1}{2}\left(\frac{\mu_i - \mu_j}{\sigma_i - \sigma_j}\right)^2} + \frac{|\mu_i - \mu_j|}{|\sigma_i - \sigma_j|} \operatorname{erf}\left(\frac{|\mu_i - \mu_j|}{\sqrt{2}|\sigma_i - \sigma_j|}\right)\right] \cdots ⑧ - \mathrm{b}$$

In the case of the Fletcher Gaussian, as the standard deviation $\sigma$ is proportional to the center frequency $f_0$, $A$ depends solely on the center frequencies $f_{0,1}$, $f_{0,2}$ and $f_{0,3}$ of the three distributions. Substituting

$$\mu_i = f_{0,i}, \ \ \sigma_i = Cf_{0,i}$$

into the above equation, we find that $D_{ij}$ is proportional to the corresponding difference in center frequencies $|f_{0,i} - f_{0,j}|$; thus, as in the case of sine waves,

$$A = 0 \ \cdots ⑧ - \mathrm{c}$$

In other words, any three-voice chords composed of Fletcher-Gaussian tones forms a degenerate Wasserstein triangle whose area is identically zero.

Calculations based on actual data yield values slightly greater than zero. This discrepancy, however, is attributable to discretization errors introduced by representing the spectra on a finite frequency grid, and is sufficiently small compared with the areas of Wasserstein triangles formed by spectra of more complex instrumental timbres.

[20] Our group's work on this issue was also made possible by Vinko's guidance during the 1980s and 1990s

[21] During a discussion with Gérard Grisey in Paris in June 1998, he remained entirely focused on the concept of 'formants' without applying sinusoidal spectral analysis to speech. We had been preparing the results of our sinusoidal decomposition of speech sounds, but Grisey passed away that autumn, and any further opportunity for discussion was lost.